\documentclass[
reprint,
showpacs,preprintnumbers,
 amsmath,amssymb,
 aps,
 prd,
]{revtex4-1}
\usepackage{graphicx}
\usepackage{dcolumn}
\usepackage{mathtools}
\usepackage{bm}
\RequirePackage[colorlinks,breaklinks,pdftitle={ATLAS draft},pdfauthor={The ATLAS Collaboration}]{hyperref}
\hypersetup{linkcolor=blue,citecolor=blue,filecolor=black,urlcolor=blue}

\def\ifb{\mbox{fb$^{-1}$}}
\def\pt{\ensuremath{p_{\rm T}}}
\newcommand{\alphab}{\alpha_b}

\newcommand{\alphaL}{\alpha_{\rm \Lambda}}

\newcommand{\lamB}{\Lambda_b^0}
\newcommand{\lamBBar}{\bar\Lambda_b^0}
\newcommand{\lam}{\Lambda^0}

\newcommand{\bd}{B^0_d}
\newcommand{\Bd}{$B^0_d$}
\newcommand{\LamBDecay}{$\Lambda_b^0\to J/\psi(\mu^+\mu^-) \Lambda^0 (p\pi^-)$}
\newcommand{\LamBDecayShort}{$\Lambda_b^0\to J/\psi \Lambda^0$}
\newcommand{\LamBDecayTitle}{$\Lambda_{\lowercase{b}}^0\to J/\psi(\mu^+\mu^-) \Lambda^0 (\lowercase{p}\pi^-)$}

\newcommand{\BdDecay}{$B_d^0\to J/\psi(\mu^+\mu^-) K_S^0 (\pi^+\pi^-)$}
\newcommand{\LamBDecayJpsimumuLambda}{$\Lambda_b^0\to J/\psi(\mu^+\mu^-) \Lambda^0$}
\newcommand{\BdDecayJpsimumuKshort}{$B_d^0\to J/\psi(\mu^+\mu^-) K_S^0$}
\newcommand{\LamDecay}{$\Lambda^0 \to p \pi^-$}
\newcommand{\LamBarDecay}{$\bar\Lambda^0 \to \bar{p} \pi^+$}

\newcommand{\JpsiDecay}{$J/\psi \to \mu^+ \mu^-$}
\newcommand{\Ks}{$K_S^0$}
\newcommand{\Jpsi}{$J/\psi$}
\newcommand{\LamB}{$\Lambda_b^0$}
\newcommand{\LamBBar}{$\bar\Lambda_b^0$}
\newcommand{\Lam}{$\Lambda^0$}
\newcommand{\LamBar}{$\bar\Lambda^0$}

\def\TeV{\ifmmode {\mathrm{\ Te\kern -0.1em V}}\else
                   \textrm{Te\kern -0.1em V}\fi}
\def\GeV{\ifmmode {\mathrm{\ Ge\kern -0.1em V}}\else
                   \textrm{Ge\kern -0.1em V}\fi}
\def\MeV{\ifmmode {\mathrm{\ Me\kern -0.1em V}}\else
                   \textrm{Me\kern -0.1em V}\fi}
\def\jpsi{\ensuremath{J/\psi}}

\usepackage{preprintcover}
\PreprintCoverPaperTitle{Measurement of the parity-violating asymmetry parameter \boldmath{$\alphab$} and the helicity amplitudes for the decay \boldmath{$\Lambda_b^0\to J/\psi\Lambda^0$} with the ATLAS detector}
\PreprintIdNumber{CERN-PH-EP-2014-034}
\PreprintCoverAbstract{A measurement of the parity-violating decay asymmetry parameter, $\alphab$, and the helicity amplitudes for the decay \LamBDecay\ is reported.
The analysis is based on 1400 \LamB\ and \LamBBar\ baryons
selected in $4.6~\ifb$ of proton--proton collision data with a center-of-mass energy of 7~TeV recorded by the ATLAS experiment at the LHC.
By combining the \LamB\ and \LamBBar\ samples under the assumption of $CP$ conservation, the value of $\alphab$ is measured to be $0.30\pm0.16 ({\rm stat})\pm 0.06 ({\rm syst})$.
This measurement provides a test of theoretical models based on perturbative QCD or heavy-quark effective theory.}
\PreprintJournalName{Physical Review D}

\begin{document}

\title{Measurement of the parity-violating asymmetry parameter \boldmath{$\alphab$} and the helicity amplitudes for the decay \boldmath{$\Lambda_b^0\to J/\psi\Lambda^0$} with the ATLAS detector}
\author{The ATLAS Collaboration}
\date{\today}

\begin{abstract}
A measurement of the parity-violating decay asymmetry parameter, $\alphab$, and the helicity amplitudes for the decay \LamBDecay\ is reported. 
The analysis is based on 1400 \LamB\ and \LamBBar\ baryons 
selected in $4.6~\ifb$ of proton--proton collision data with a center-of-mass energy of 7~TeV recorded by the ATLAS experiment at the LHC.
By combining the \LamB\ and \LamBBar\ samples under the assumption of $CP$ conservation, the value of $\alphab$ is measured to be $0.30\pm0.16 ({\rm stat})\pm 0.06 ({\rm syst})$. 
This measurement provides a test of theoretical models based on perturbative QCD or heavy-quark effective theory.
\end{abstract}

\pacs{14.20.Mr}
\maketitle

\section{Introduction}
Parity violation, a well-known feature of weak interactions~\cite{Lee:1956qn,Wu:1957my,Garwin:1957hc,Eisler:1957ih}, is exhibited in its maximal form in decays of muons and $\tau$ leptons. However, in weak decays of hadrons, it is not maximal and depends on the hadron's constituents because of the presence of strongly bound spectator quarks.
For example, the process \LamDecay\ has a parity-violating decay asymmetry parameter, $\alphaL$, of over 0.6~\cite{Cronin:1963zb}.
The decay asymmetry parameter $\alpha$ enters into the angular distribution of any two-body spin 1/2 particle decay as follows:
\begin{equation}
w(\cos\theta) = \frac{1}{2}(1+\alpha P \cos\theta),
\end{equation}
where $P$ is the polarization of the particle and $\theta$ is defined as the angle between the polarization vector and the direction of the decay product in the particle's rest frame.
The strong interaction effects in the hadron decays are non-perturbative, which makes it very difficult to predict the value of $\alpha$, at least for light hadrons such as \Lam. However, in the case of heavy baryons, such as \LamB, the energy release in the decay of the $b$-quark is large enough that the use of the factorization theorem and perturbative QCD (pQCD) seems justified to compute the effects of the strongly coupled spectator quarks, making theoretical predictions possible.

Several models have been employed to predict the value of the parity-violating decay asymmetry parameter $\alphab$ for the weak decay $\Lambda_b^0\to J/\psi\Lambda^0$.
Various quark models are used to calculate the form factors in the factorization approximation (FA)~\cite{Cheng:1996cs,Fayyazuddin:1998ap,Mohanta:1998iu,Wei:2009np,Gutsche:2013oea} and the predictions of $\alphab$ generally lie in the range from $-0.2$ to $-0.1$.
In Ref.~\cite{Chou:2001bn}, the \LamBDecay\ decay process is factorized into parts calculable in pQCD and universal hadron distribution amplitudes, so both the factorizable and non-factorizable contributions in the FA are included. The value of $\alphab$ is predicted to be in the range from $-0.17$ to $-0.14$.
However, a calculation based on heavy-quark effective theory (HQET)~\cite{Leitner:2004dt,Ajaltouni:2004zu} predicts a value 0.78.

Recently, the LHCb experiment reported a measurement of $\alphab=0.05 \pm 0.17({\rm stat}) \pm 0.07({\rm syst})$~\cite{Aaij:2013oxa}.
This paper provides a measurement of comparable precision using 4.6~fb$^{-1}$ $pp$ collision data recorded by the ATLAS detector with a center-of-mass energy of 7~TeV.
\section{The \boldmath{\LamBDecayTitle}\ decay}
\label{sec:theory}
\begin{figure}[bt]
    \begin{center}
      \includegraphics[width=0.75\linewidth]{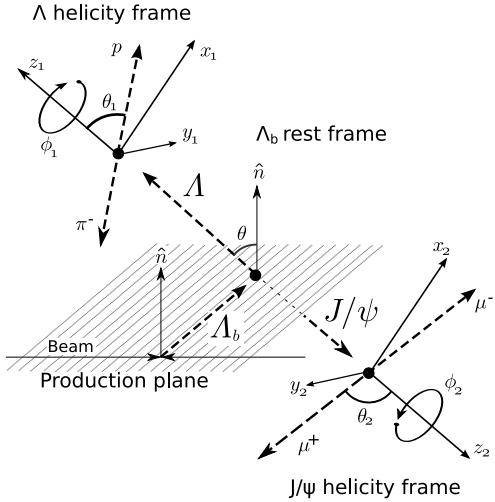}
    \end{center}
    \caption{The decay angles, as defined in the text.}
    \label{fig:angles}
  \end{figure}

Because of parity conservation, \LamB\ produced by the strong interaction, which is the dominant production mechanism, can be polarized only in a direction perpendicular to the \LamB\ production plane, $\hat{n}$~\cite{Ajaltouni:2004zu,Hrivnac:1994jx}. The vector $\hat{n}$ is chosen to point in the direction of the cross product of the beam direction and the \LamB\ momentum.
Since the LHC collides proton beams traveling in opposite directions, either beam direction could be used. 
This analysis uses the positive $z$-axis direction of 
the ATLAS coordinate system~\footnote{
ATLAS uses a right-handed coordinate system with its origin at the nominal interaction point (IP) in the center of the detector and the $z$-axis along the beam pipe. The $x$-axis points from the IP to the center of the LHC ring, and the $y$-axis points upward. Cylindrical coordinates $(r,\Phi)$ are used in the transverse plane, $\Phi$ being the azimuthal angle around the beam pipe. The pseudorapidity is defined in terms of the polar angle $\Theta$ as $\eta=-\ln\tan(\Theta/2)$.
} 
for the \LamB\ candidates and the negative $z$-axis for \LamBBar\ candidates (to preserve symmetry between \Lam\ and \LamBar\ given by the orientation of the ATLAS magnetic field). 
The definition of the decay angles is shown in Fig.~\ref{fig:angles}. The angle $\theta$ is the polar angle of the \Lam\ momentum measured from the normal direction $\hat n$ in the \LamB\ rest frame. The corresponding azimuthal angle, $\phi$, is of no interest in this analysis and therefore is not labeled in the figure. 
The angles $\theta_1$ ($\theta_2$) and $\phi_1$ ($\phi_2$) are the polar and azimuthal angles of the proton ($\mu^+$) in the \Lam\ (\Jpsi) rest frame with respect to the \Lam\ (\Jpsi) direction in the \LamB\ rest frame.
 The azimuthal angles, $\phi_1$ and $\phi_2$, are measured in the right-handed coordinate systems of the rest frames of \Lam\ and \Jpsi, $(x_1,y_1,z_1)$ and $(x_2,y_2,z_2)$, respectively. The $z_{1,2}$-axes are aligned with the direction of \Lam\ and \Jpsi, respectively, and the $x_{1,2}$-axes lie in the plane containing $\hat n$ and the \Lam\ or \Jpsi\ momenta. With this definition, the sum $\phi_1+\phi_2$ gives the angle between the \Lam\ and \Jpsi\ decay planes.

Taking $\lambda_{\Lambda}$ and $\lambda_{J/\psi}$ to represent the helicity of the \Lam\ and the \Jpsi, the decay \LamBDecayShort\ can be described by four helicity amplitudes $A(\lambda_{\Lambda}, \lambda_{J/\psi})$:
$a_+ \equiv A(1/2, 0)$, $a_-\equiv A(-1/2, 0)$, $b_+\equiv A(-1/2, -1)$ and $b_-\equiv A(1/2, 1)$, which are normalized to unity:
\begin{equation}
 |a_+|^2+|a_-|^2+|b_+|^2+|b_-|^2 = 1.
 \label{eqn_helicity_normliaztion}
\end{equation}
The full angular probability density function (PDF) of the decay angles $\Omega=(\theta,\phi,\theta_1,\phi_1,\theta_2,\phi_2)$ is~\cite{Lednicky:1985zx, Bialas:1992ny, Hrivnac:1994jx}:
\begin{equation}
 w(\Omega, \vec{A},P)=\frac{1}{(4\pi)^3}\sum_{i=0}^{19}f_{1i}(\vec{A})f_{2i}(P,\alphaL)F_{i}(\Omega),
\label{eqn:www}
\end{equation}
with the 20 terms $f_{1i}$, $f_{2i}$ and $F_i$ listed in Table~\ref{tab:pdf}. $\vec{A}$ represents the four helicity amplitudes and $P$ is the polarization of \LamB.
Under the assumption of $CP$ conservation in \LamDecay\ and \LamBarDecay\ decays, $\alpha_{\bar\Lambda} = -\alphaL = -0.642\pm0.013$ is used in this analysis, because the value $\alphaL= 0.642\pm0.013$ is measured with better precision than its counterpart $\alpha_{\bar\Lambda} = -0.71 \pm 0.08$~\cite{PDG}.
The $F_i(\Omega)$ are orthogonal functions of the decay angles.
  \begin{table*}[tb]
  \caption{The coefficients $f_{1i}^{}$, $f_{2i}^{}$ and $F_i^{}$ of the probability density
    function in Eq.~(\ref{eqn:www})~\cite{Hrivnac:1994jx}.}
  \label{tab:pdf}
  \centering
    \begin{tabular}{cccc}
      \hline\hline
      $i$&$f_{1i}^{}$&$f_{2i}^{}$&$F_i^{}$\\
      \hline
      0 &$a_+^{} a_+^\ast + a_-^{} a_-^\ast + b_+^{} b_+^\ast +b_-^{} b_-^\ast$  & $1$&1\\
      \hline
      1 &$a_+^{} a_+^\ast - a_-^{} a_-^\ast + b_+^{} b_+^\ast - b_-^{} b_-^\ast$ & $P$ &$\cos\theta$\\
      2 &$a_+^{} a_+^\ast - a_-^{} a_-^\ast - b_+^{} b_+^\ast + b_-^{} b_-^\ast$&$\alpha_\Lambda^{} $&$\cos\theta_1^{}$\\
      3 &$a_+^{} a_+^\ast + a_-^{} a_-^\ast - b_+^{} b_+^\ast - b_-^{} b_-^\ast$&$P\,\alpha_\Lambda^{} $&$\cos\theta \, \cos\theta_1^{} $\\
      4 &$-a_+^{} a_+^\ast - a_-^{} a_-^\ast + \frac{1}{2} b_+^{} b_+^\ast + \frac{1}{2} b_-^{} b_-^\ast$&$1$&$\frac{1}{2} \, (3\cos^2\theta_2^{}-1)$\\
      5 &$-a_+^{} a_+^\ast + a_-^{} a_-^\ast + \frac{1}{2} b_+^{} b_+^\ast - \frac{1}{2} b_-^{} b_-^\ast$&$P$&$ \frac{1}{2} \, (3\cos^2\theta_2^{}-1) \,\cos\theta$\\
      6 &$-a_+^{} a_+^\ast + a_-^{} a_-^\ast - \frac{1}{2}b_+^{} b_+^\ast+\frac{1}{2}b_-^{} b_-^\ast$&$\alpha_\Lambda^{}$&$ \frac{1}{2} \, (3\cos^2\theta_2^{}-1) \,\cos\theta_1^{}$\\
      7 &$- a_+^{} a_+^\ast - a_-^{} a_-^\ast - \frac{1}{2} b_+^{} b_+^\ast - \frac{1}{2}b_-^{} b_-^\ast$ & $P\,\alpha_\Lambda^{}$&$ \frac{1}{2} \, (3\cos^2\theta_2^{}-1) \,\cos\theta\cos\theta_1^{}$\\
      \hline
      8 &$-3\mathrm{Re}(a_+^{}a_-^\ast)$&$P\,\alpha_\Lambda^{}$ &$\sin\theta\,\sin\theta_1^{}\,\sin^2\theta_2^{}\,\cos\phi_1^{}$\\
      9 &$3\mathrm{Im}(a_+^{}a_-^\ast)$&$P\,\alpha_\Lambda^{}$& $\sin\theta\,\sin\theta_1^{}\,\sin^2\theta_2^{}\,\sin\phi_1^{}$\\
      10 &$-\frac{3}{2}\mathrm{Re}(b_-^{} b_+^\ast)$&$P\,\alpha_\Lambda^{}$ &$\sin\theta\,\sin\theta_1^{}\,\sin^2\theta_2^{}\,\cos(\phi_1^{}+2\,\phi_2^{})$\\
      11 &$\frac{3}{2}\mathrm{Im}(b_-^{}b_+^\ast)$&$P\,\alpha_\Lambda^{}$ &$\sin\theta\,\sin\theta_1^{}\,\sin^2\theta_2^{}\,\sin(\phi_1^{}+2\,\phi_2^{})$\\
      \hline
      12 &$-\frac{3}{\sqrt{2}}\mathrm{Re}(b_-^{}a_+^\ast+a_-^{}b_+^\ast)$&$P\,\alpha_\Lambda^{}$ &$\sin\theta\,\cos\theta_1^{}\,\sin\theta_2^{}\,\cos\theta_2^{}\,\cos\phi_2^{}$\\
      13 &$\frac{3}{\sqrt{2}}\mathrm{Im}(b_-^{}a_+^\ast + a_-^{}b_+^\ast)$ &$P\,\alpha_\Lambda^{}$ &$\sin\theta\,\cos\theta_1^{}\,\sin\theta_2^{}\,\cos\theta_2^{}\,\sin\phi_2^{}$\\
      14 &$-\frac{3}{\sqrt{2}}\mathrm{Re}(b_-^{}a_-^\ast+a_+^{}b_+^\ast)$  &$P\,\alpha_\Lambda^{}$ &$\cos\theta\,\sin\theta_1^{}\,\sin\theta_2^{}\,\cos\theta_2^{}\,\cos(\phi_1^{}+\phi_2^{})$\\
      15 &$\frac{3}{\sqrt{2}}\mathrm{Im}(b_-^{}a_-^\ast+a_+^{}b_+^\ast)$  &$P\,\alpha_\Lambda^{}$ &$\cos\theta\,\sin\theta_1^{}\,\sin\theta_2^{}\,\cos\theta_2^{}\, \sin(\phi_1^{}+\phi_2^{})$\\
      \hline
      16 &$\frac{3}{\sqrt{2}}\mathrm{Re}(a_-^{}b_+^\ast-b_-^{}a_+^\ast)$&$P$ &$\sin\theta\,\sin\theta_2^{}\,\cos\theta_2^{}\,\cos\phi_2^{}$\\
      17 &$-\frac{3}{\sqrt{2}}\mathrm{Im}(a_-^{}b_+^\ast-b_-^{}a_+^\ast)$&$P$ &$\sin\theta\,\sin\theta_2^{}\,\cos\theta_2^{}\,\sin\phi_2^{}$\\
      18 &$\frac{3}{\sqrt{2}}\mathrm{Re}(b_-^{}a_-^\ast - a_+^{}b_+^\ast)$ &$\alpha_\Lambda^{}$&$\sin\theta_1^{}\,\sin\theta_2^{}\,\cos\theta_2^{}\,\cos(\phi_1^{}+\phi_2^{})$\\
      19 &$-\frac{3}{\sqrt{2}}\mathrm{Im}(b_-^{}a_-^\ast-a_+^{}b_+^\ast)$ &$\alpha_\Lambda^{}$&$\sin\theta_1^{}\,\sin\theta_2^{}\,\cos\theta_2^{}\, \sin(\phi_1^{}+\phi_2^{})$\\
      \hline
      \hline
    \end{tabular}
  \end{table*}

The $\alphab$ parameter is related to the helicity amplitudes as follows~\cite{Hrivnac:1994jx}:
\begin{equation}
 \alphab = |a_+|^2-|a_-|^2+|b_+|^2-|b_-|^2.
\label{eqn_alphab_def}
\end{equation}
There are nine unknown real parameters in the PDF (Eq.~(\ref{eqn:www})): four complex helicity amplitudes, $a_+=|a_+|e^{i\rho_{+}}$, $a_-=|a_-|e^{i\rho_{-}}$, $b_+=|b_+|e^{i\omega_{+}}$, $b_-=|b_-|e^{i\omega_{-}}$, each with a magnitude and a phase, and the polarization $P$. However, only six out of the eight helicity amplitude parameters are independent, taking into account the normalization constraint (Eq.~(\ref{eqn_helicity_normliaztion}))
and, due to the arbitrary value of the common phase, only differences between the four phases are relevant.

The angular PDF is further simplified due to the symmetry of the initial state at a $pp$ collider.
Since the arbitrary choice of the beam direction cannot bear on the physics result, the polarization must be an odd function of the \LamB\ pseudorapidity: $P(\pt,\eta) = -P(\pt, -\eta)$.
Therefore, for a sample of \LamB\ produced over a symmetric interval in pseudorapidity, which is satisfied in the ATLAS detector, the average polarization must be zero.
As a result, only six terms in Table~\ref{tab:pdf} which are not dependent on $P$ are retained in the PDF and they depend only on five independent parameters: three magnitudes of the helicity amplitudes and two relative phases. 
The remaining phase cannot be resolved with a zero-polarization sample, but
$\alphab$ can be determined from the magnitudes of the helicity amplitudes as in Eq.~(\ref{eqn_alphab_def}).
The following choice of the fit model parameterization is found to have only a small correlation of uncertainties and is used in this analysis:
\begin{align}
\alphab &= |a_+|^2-|a_-|^2+|b_+|^2-|b_-|^2, \nonumber \\
k_+ &= \frac{|a_+|}{\sqrt{|a_+|^2+|b_+|^2}}, \nonumber \\
k_- &= \frac{|b_-|}{\sqrt{|a_-|^2+|b_-|^2}}, \nonumber \\
\Delta_+ &= \rho_+-\omega_+, \nonumber \\
\Delta_- &= \rho_--\omega_-.
\label{eqn:paras}
\end{align}
Table~\ref{tab:pdf2} shows the explicit dependence of the $f_{1i}$ functions on the chosen parameters.

  \begin{table*}[tb]
  \centering
    \caption{The coefficients $f_{1i}^{}$ of the remaining six terms of the simplified PDF expressed using the five free parameters defined in Eq.~(\ref{eqn:paras}).}
    \label{tab:pdf2}
   \begin{tabular}{cl}
      \hline\hline
      $i$&$f_{1i}^{}$\\
      \hline
      0 &$1$ \\
      2 &$(k^2_++k^2_--1)+\alphab(k^2_+-k^2_-)$ \\
      4 &$\frac{1}{4}[(3k^2_--3k^2_+-1)+3\alphab(1-k^2_--k^2_+)]$ \\
      6 &$-\frac{1}{4}[(k^2_++k^2_--1)+\alphab(3+k^2_+-k^2_-)]$ \\
      18 &$\frac{3}{\sqrt{2}}[\frac{1-\alphab}{2}\sqrt{k^2_-(1-k^2_-)}\cos(-\Delta_-)-\frac{1+\alphab}{2}\sqrt{k^2_+(1-k^2_+)}\cos(\Delta_+)]$ \\
      19 &$-\frac{3}{\sqrt{2}}[\frac{1-\alphab}{2}\sqrt{k^2_-(1-k^2_-)}\sin(-\Delta_-)-\frac{1+\alphab}{2}\sqrt{k^2_+(1-k^2_+)}\sin(\Delta_+)]$ \\
      \hline \hline
    \end{tabular}
  \end{table*}

If $CP$ is conserved, the PDFs of the \LamB\ and \LamBBar\ decays have exactly the same form.
Therefore, assuming $CP$ conservation, the \LamB\ and \LamBBar\ samples are combined to measure $\alphab$ and the helicity amplitudes.
\section{Data samples and trigger selection}
\label{section:data}
ATLAS~\cite{ATLAS} covers nearly the entire solid angle around the interaction point with layers of tracking detectors, 
calorimeters, and muon chambers. This analysis uses two sub-systems: the inner detector (ID) and the muon spectrometer (MS). 
The ID consists of three types of detectors: a silicon pixel detector (Pixel), a silicon micro-strip detector (SCT) and a transition radiation tracker (TRT). 
These detectors are surrounded by a thin superconducting solenoid providing a 2~T axial magnetic field.
The MS measures the deflection of muons in a magnetic field produced by three large superconducting air-core toroid systems, each with eight superconducting coils, and it consists of four sub-detectors.
Monitored drift tube chambers and cathode strip chambers are used for precision muon measurements, while
resistive plate chambers (RPCs) and thin gap chambers (TGCs) are used by the muon trigger system.
The MS and ID provide a pseudorapidity coverage up to $|\eta|=2.5$.
Tracks reconstructed in the ID with $\pt>400\MeV$ are used in this analysis.

This analysis uses 7~TeV collision data collected in 2011 with single-muon triggers and the dimuon triggers used to select \JpsiDecay. The corresponding integrated luminosity is $4.6~\ifb$~\cite{Aad:2013ucp}. The ATLAS trigger system \cite{Aad:2012xs} has three levels: the hardware-based Level-1 trigger and the two-stage High-Level Trigger (HLT). 
At Level-1, the muon trigger uses RPCs and TGCs to search for patterns of hits corresponding to muons passing different \pt\ thresholds. 
Regions of interest around these Level-1 hit patterns then serve as seeds for the HLT muon reconstruction. 
When the rate from the low-\pt\ muon triggers exceeded the allotted trigger bandwidth,
prescale factors were applied to reduce the output rate. The transverse momentum threshold for unprescaled single-muon triggers was 18~GeV. The \JpsiDecay\ triggers are dimuon triggers that require the muons to have opposite charge and the dimuon mass to be in the interval $2.5 < m_{\mu\mu} < 4.3\GeV$. 
Most of the sample was collected by the \JpsiDecay\ trigger with a \pt\ threshold of 4~GeV applied to both muons. This is the lowest \pt\ threshold trigger unprescaled in the 2011 data-taking period. 
\section{Monte Carlo samples}
A Monte Carlo (MC) sample of signal events is used to study the efficiency and acceptance of the detector. Inclusive inelastic events are generated using the {\sc Pythia} 6.4 MC generator~\cite{Pythia} and filtered such that each event contains a signal decay, \LamBDecayJpsimumuLambda, with the muons having transverse momenta above 2.5~GeV. In addition to the \LamB\ MC sample, \BdDecayJpsimumuKshort\ and $b\bar{b}\to J/\psi(\mu^+\mu^-)+X$ MC samples are also generated with the same generator-level muon cuts in order to optimize the selection cuts and understand the
sources of background. The MC events are passed through the ATLAS simulation and reconstruction software~\cite{Aad:2010ah} based on the {\sc Geant}~4~\cite{Agostinelli:2002hh} package for the detector simulation. 
The MC simulation and reconstruction software is configured to reproduce the detector conditions during data taking.
\section{Reconstruction and signal selection}
\subsection{Muon Reconstruction}
\label{sec:muon}
Two types of muons are used in the analysis, known as tagged muons and combined muons~\cite{Aad:2011rr}. A charged-particle track reconstructed in the MS is matched to one reconstructed in the ID to form a combined muon. The pseudorapidity coverage of combined muons is $|\eta|<2.5$. Tagged muons, consisting of tracks reconstructed in the ID and matched to patterns of hits in the MS, cover the pseudorapidity range $|\eta|<2.2$ and contribute to the muon reconstruction efficiency in the low-\pt\ range. 
Although both the ID and the MS provide a momentum measurement separately, only the ID measurement is used because of its better resolution in the \pt\ range relevant for this analysis, and the MS is used only to identify muons.
The reconstructed muon tracks are required to have a sufficient number of hits in the Pixel, SCT and TRT detectors to ensure accurate ID measurements.

\subsection{\boldmath{\Jpsi} and \boldmath{\Lam} Preselection}

The decay \LamBDecay\ has a cascade topology, as the \Jpsi\ decays instantly at the same point as the \LamB\ (forming a secondary vertex) while \Lam\ lives long enough to form a displaced tertiary vertex.

The \Jpsi\ candidates are selected by fitting dimuon pairs to a common vertex~\cite{VKalVrt} and requiring that their invariant mass lies in the range $2.8 < m_{\mu\mu} < 3.4 \GeV$. The dihadron pairs are also fitted to a common vertex and accepted as \Lam\ candidates if the invariant mass is in the range $1.08 < m_{p\pi} < 1.15 \GeV$. The tracks used for the primary vertex reconstruction are excluded from the \Lam\ vertex fit to reduce the large combinatorial background.
The masses of a proton and a pion are assigned to the tracks when the invariant mass is calculated; $p\pi^-$ and $\bar p\pi^+$ combinations are considered so that both the \Lam\ and \LamBar\ candidates are accepted.

\subsection{Reconstruction of \boldmath{\LamBDecay}}
\label{section:lamb}

The preselected muon and hadron track pairs are then refitted with a constraint to the \LamBDecay\ topology. 
The muons are required to intersect at a single vertex and their invariant mass is constrained to the mass of the \Jpsi, $m_{J/\psi} = 3096.9$~MeV~\cite{PDG}. The two hadron tracks are forced to intersect in a second vertex and their invariant mass is fixed to the mass of the \Lam, $m_{\Lambda^0} = 1115.7$~MeV \cite{PDG}. The combined momentum direction of the refitted \Lam\ track pair is constrained to point to the dimuon vertex. Two mass hypotheses are considered: the first assigns the proton mass to the positive track and the pion mass to the negative track, and the second hypothesis makes the opposite mass assignment. These hypotheses correspond to \LamB\ and \LamBBar\ decays, respectively. The fit is performed on all four tracks simultaneously, taking into account the constraints described above~\cite{VKalVrt}.
The quality of the fit is characterized by the value of $\chi^2$ divided by the number of degree of freedom, $N_{\rm dof}$.
Furthermore, for each track quadruplet that can be successfully fitted to the \LamB\ decay topology, a fit to the \BdDecay\ decay topology is attempted (i.e. the pion mass is assigned to the hadron tracks and the dihadron mass is constrained to the mass of \Ks, $m_{K_S} = 497.6$~MeV~\cite{PDG}). The \Bd\ fit is needed to identify possible \Bd\ decays misidentified as \LamB.

The fitted $\lamB$ are further required to pass following selection cuts (see Ref.~\cite{Aad:2012bpa} for details):
\begin{itemize}
  \item The fit quality $\chi^2/N_{\rm dof} < 3$.
  \item The transverse momentum of the refitted \Lam, $p_{\rm T,\lam} > 3.5$~GeV.
  \item The transverse decay length of the refitted \Lam\ vertex measured from the \LamB\ vertex, $L_{xy,\lam} > 10$~mm.
  \item If the four tracks forming a \LamB\ candidate also result in an acceptable \Bd\ fit, the candidate must have a larger cumulative $\chi^2$ probability for the \LamB\ fit: $\mathcal{P}_{\lamB} > \mathcal{P}_{\bd}$.
  \item The reconstructed \LamB\ proper decay time~\cite{Aad:2012bpa}, $\tau>0.35$~ps.
\end{itemize}
\begin{figure}[tb]
\includegraphics[width=0.45\textwidth]{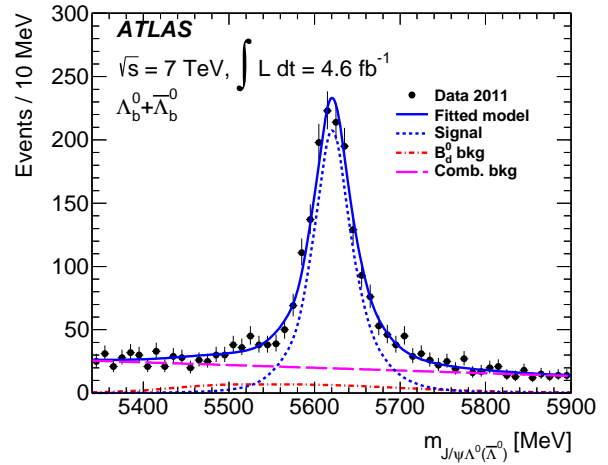}
\caption{The reconstructed mass of \LamB\ and \LamBBar\ candidates, fitted with a three-component PDF (blue solid curve) consisting of signal (blue dashed curve), combinatorial (magenta long-dashed straight line) and \Bd\ background (red dot-dashed curve, bottom).}
\label{fig_mass_fit}
\end{figure}

Figure~\ref{fig_mass_fit} shows the invariant mass distribution of events passing these selection cuts in the range from 5340~MeV to 5900~MeV.
There is no track quadruplet simultaneously satisfying both the $\lamB$ and $\lamBBar$ hypotheses.
Background events can be divided into two categories: the combinatorial background and the peaking background. The combinatorial background consists of real or fake \Jpsi\ and \Lam\ candidates randomly combined to create a \LamB-like topology.
This is the main component of the background, whose mass distribution is non-resonant and assumed to be linear in the vicinity of the \LamB\ mass.
The peaking background is due to residual \BdDecay\ decays passing the requirement $\mathcal{P}_{\lamB} > \mathcal{P}_{\bd}$.
The invariant mass distribution is fitted with a three-component PDF to estimate the number of signal, combinatorial background and \Bd\ background events.
The shapes of the \LamB\ signal component and the \Bd\ background are modeled using one-dimensional Gaussian-kernel estimation PDFs \cite{Cranmer:2000du} of the MC events. The Gaussian-kernel estimators are non-parametric PDFs describing the shape of the invariant mass distribution of the MC candidates (i.e. MC templates). 
The advantage of using MC templates is that they accurately describe the non-Gaussian tails of the \LamB\ peak as well as the asymmetry of the \Bd\ background, which is important in correctly estimating the number of events in the fit.
The effect of possible mis-modeling of the shape of $m_{J/\psi\lam}$ in the signal MC sample is discussed in Sec.~\ref{sec_syst}.
The combinatorial background is parameterized by a first-order polynomial.
An extended binned maximum likelihood fit~\cite{Verkerke:2003ir} is performed with the number of events corresponding to each PDF component ($N_{\rm sig}$, $N_{\rm Comb}$, and $N_{\bd}$) and the slope of the linear background PDF as free parameters.

The numbers of events extracted by the invariant mass fit are summarized in Table~\ref{tab:m_fit}.
A mass window around the nominal \LamB\ mass~\cite{PDG}, $5560 < m_{J/\psi\lam} < 5680\MeV$, is defined as the signal region (SR) for this measurement.
In the SR, the number of \Bd\ events is nearly one fourth of the total number of background events, and it has a large relative uncertainty due to its small size and the broad distribution of the \Bd\ peak.
\begin{table}[tb]
 \centering
\caption{The numbers of signal candidates, $N_{\rm sig}$, combinatorial background, $N_{\rm Comb}$, and \Bd\ background candidates, $N_{\bd}$, extracted by the extended binned maximum likelihood fit in the mass range from 5340~MeV to 5900~MeV. The number of events from each component in the SR mass window is given by scaling the values from the fit.}
\label{tab:m_fit}
\begin{tabular}{lcc}
\hline \hline
Parameter & $[5340,5900]\MeV$ & $[5560,5680]\MeV$ \\
 \hline
 $N_{\rm sig}$ & $1400 \pm 50$ & $1240 \pm 40$\\
 $N_{\rm Comb}$     & $1090 \pm 80$ & $\phantom{1}234 \pm 16$\\
 $N_{\bd}$       & $\phantom{1}210 \pm 90$ & $\phantom{12}73 \pm 30$\\
\hline \hline
\end{tabular}
\end{table}
\section{Parameter extraction}
\subsection{Least squares fit}
The average values of the angular distributions $F_i(\Omega)$ defined in Table~\ref{tab:pdf}:
\begin{equation}
  \langle F_i \rangle = \frac{1}{N^{\rm data}}\sum\limits_{n=1}^{N^{\rm data}} F_i(\Omega_n)
\end{equation}
are used to extract the helicity parameters.
By definition, $\langle F_0\rangle$ is identical to one.

The expected values of $\langle F_i \rangle$ depend on the helicity parameters $\vec{A}$ and can be obtained by convolving these functions with the PDF (Eq.~(\ref{eqn:www})) and integrating over the full angular range:
\begin{equation}
\langle F_i \rangle^{\rm expected} = \sum_j f_{1i}(\vec{A})f_{2i}(\alphaL) C_{ij},
\label{eqn_mm1}
\end{equation}
with
\begin{equation}
\mathrm{C}_{ij} = \frac{1}{(4\pi)^3}\iint F_i(\Omega') T(\Omega',\Omega) F_j(\Omega) \,\mathrm{d}\Omega'\mathrm{d}\Omega,
\label{eqn:C_matrix}
\end{equation}
where $\Omega$ stands for the true decay angles and $\Omega'$ for the measured ones.
The acceptance, efficiency and resolution of the detector are represented by $T(\Omega',\Omega)$.
These detector effects are encoded in the matrix $\mathbf{C}$, whose elements 
do not depend on the helicity parameters, $\vec{A}$.

Ideally, the helicity amplitude parameters can be calculated by solving the system of five equations with five parameters:
\begin{equation}
 \langle F_i\rangle^{\rm expected} = \langle F_i \rangle, \mbox{ \quad for $i=2,4,6,18,$ and 19.} \label{eqn:eqn_solution}
\end{equation}
However, with the measured values of $\langle F_i\rangle$ in current data, Eq.~(\ref{eqn:eqn_solution}) has no solution with real parameters.
Therefore, the set of real parameters that are statistically closest to the exact solution is found by minimizing the $\chi^2$ function with respect to the five real parameters:
\begin{equation}
\chi^2 = \sum_{i}\sum_{j} (\langle F_i \rangle^{\rm expected} - \langle F_i \rangle)V_{ij}^{-1}(\langle F_j\rangle^{\rm expected} - \langle F_j\rangle),
\label{eqn_chi2}
\end{equation}
where $i,j=2,4,6,18$ and 19, and $\mathbf{V}$ is the covariance matrix of the measured $\langle F_i \rangle$ values. The correlations between the five averages are accounted for by the covariance matrix.

\subsection{Background subtraction}
As the combinatorial background can be described by the linear function, 
its contribution to the measured $\langle F_i \rangle$ values can be estimated by using events in the invariant mass sidebands. Two mass windows define the sidebands: $5400<m_{J/\psi\lam}<5520\MeV$ is chosen as the left sideband and $5720<m_{J/\psi\lam}<5840\MeV$ as the right one. The background contribution to the $\langle F_i \rangle$ values in the signal region is estimated as an average of the values in the two sidebands and is subtracted from the measured value of $\langle F_i \rangle$.

The similarity of the left and right sidebands can be verified by comparing the $F_i$ distributions. Figure~\ref{fig_F_sb} shows that the distributions for $F_i$ are similar in the two sidebands while the distributions in the signal region are different. The only significant difference between the occupancy of the two sidebands is when the value of $F_6$ is close to zero and is due to \Bd\ background.
\begin{figure*}[tb]
\includegraphics[width=0.49\textwidth]{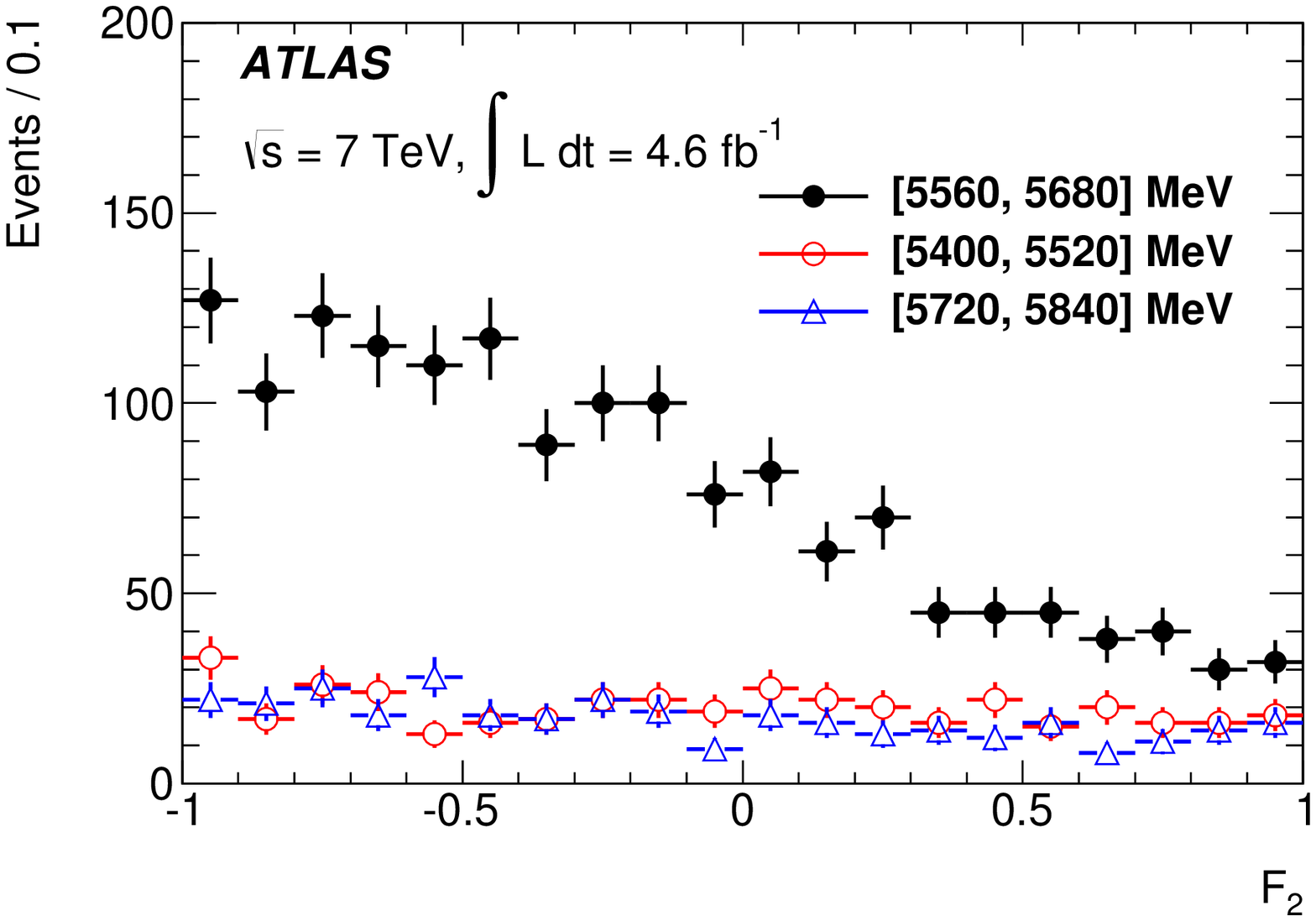}
\includegraphics[width=0.49\textwidth]{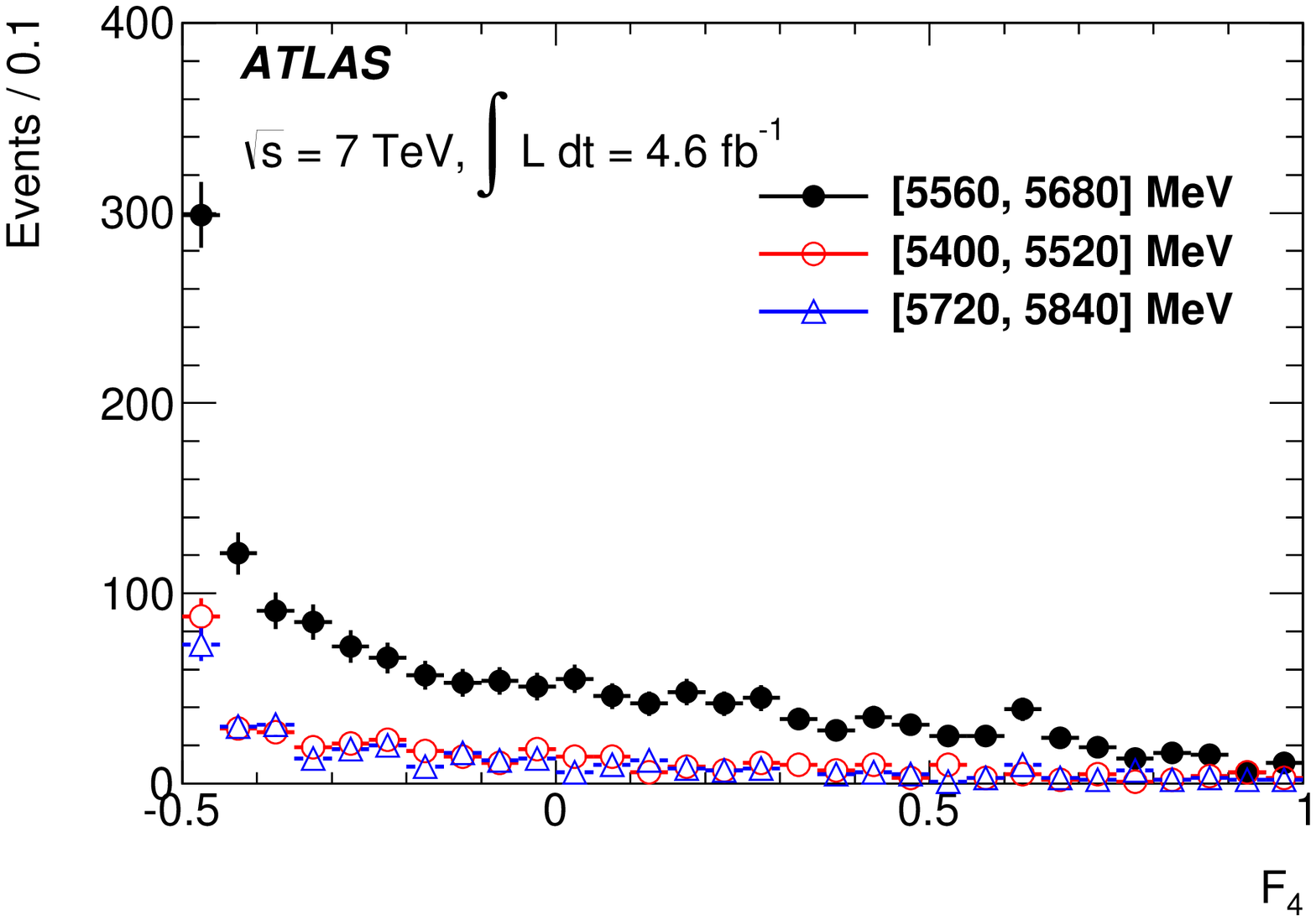}
\includegraphics[width=0.49\textwidth]{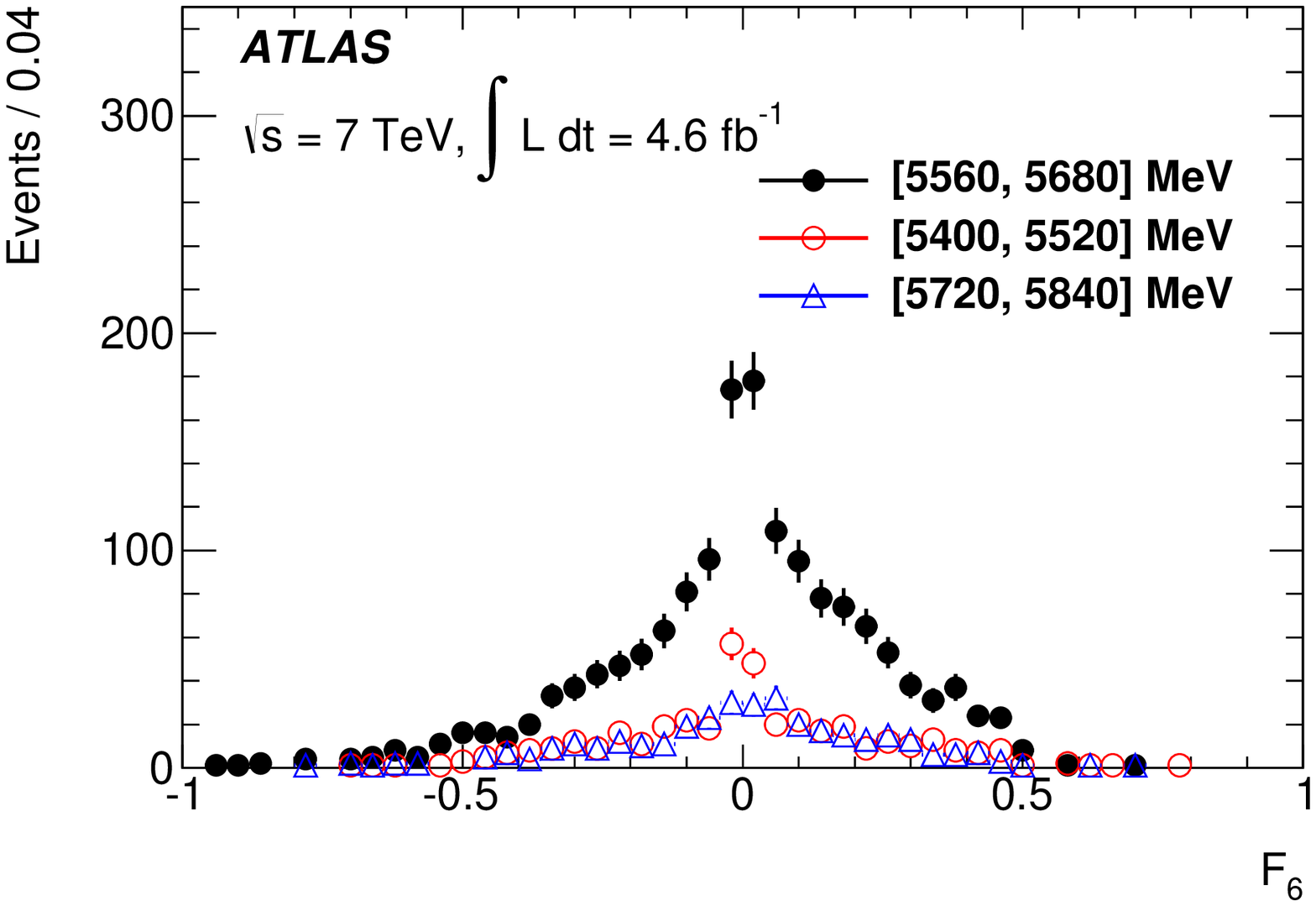}
\includegraphics[width=0.49\textwidth]{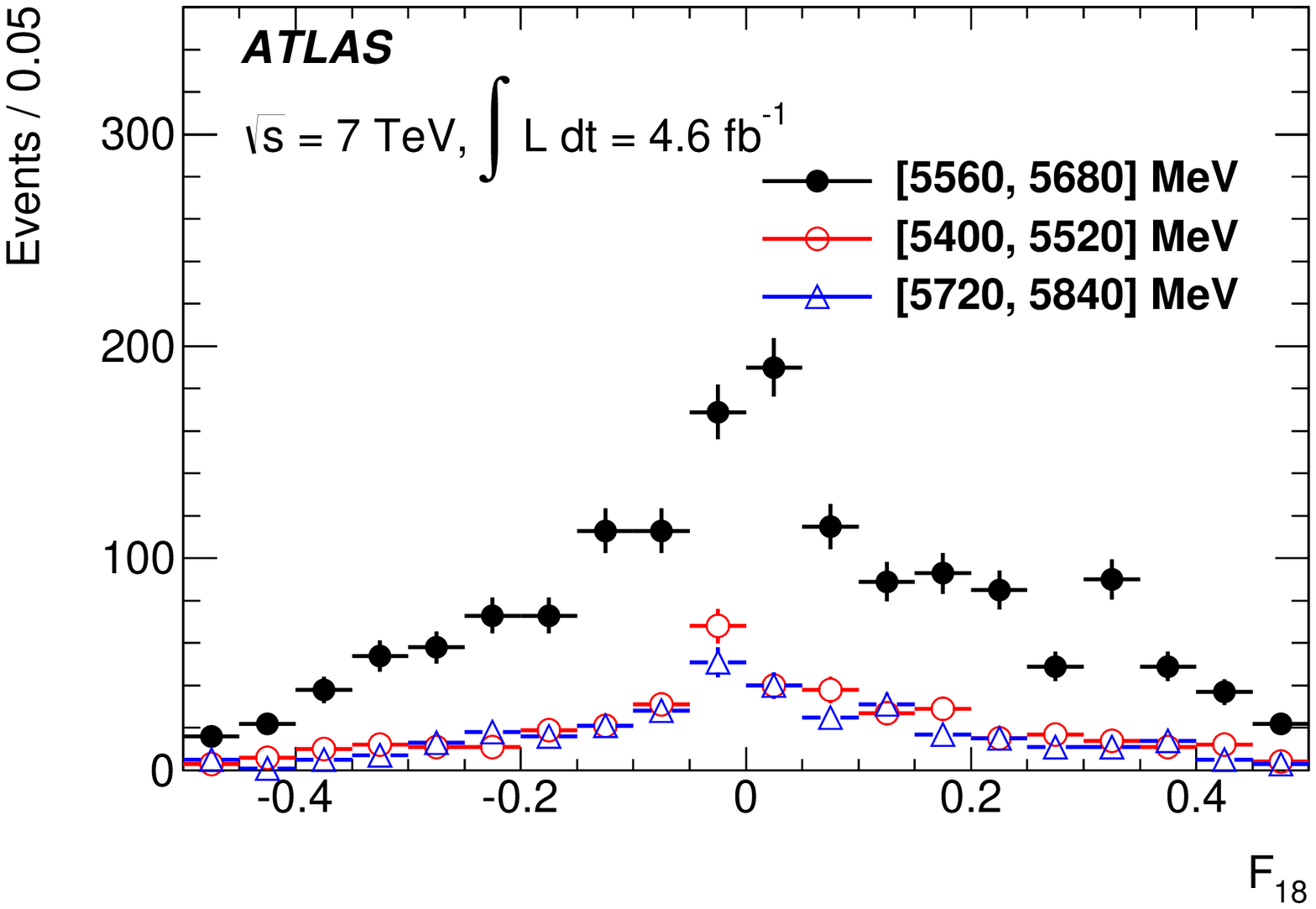}
\includegraphics[width=0.49\textwidth]{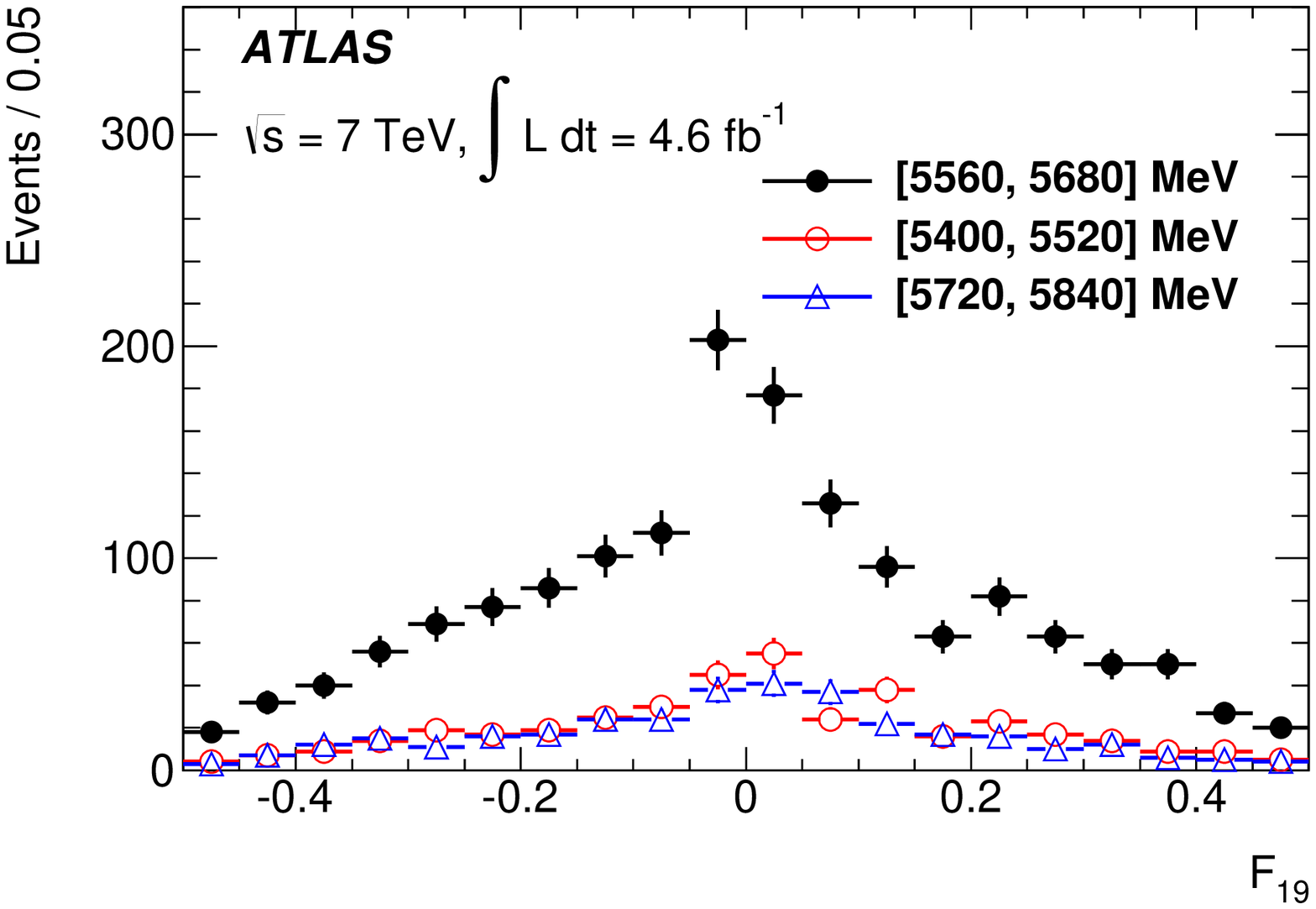}
\caption{The $F_i$ ($i=2,4,6,18,19$) distribution for events in the sidebands, together with the distribution for events in the signal region.}
\label{fig_F_sb}
\end{figure*}

The \Bd\ MC sample, together with the estimated number of \Bd\ events (Sec.~\ref{section:lamb}), is used to calculate the contribution of the \Bd\ events to the averaged $\langle F_i \rangle$ values and the estimated contribution is subtracted.

\subsection{Detector effects correction}
In the case of an ideal detector, there are no acceptance and resolution effects, i.e. $T(\Omega',\Omega)=\delta(\Omega',\Omega)$, where $\delta(\Omega',\Omega)$ is the Dirac delta function. In this case, $\mathbf{C}$ is a simple diagonal matrix $\mathbf{D}$ with elements
\begin{equation}
 \mathrm{D}_{ij} = \frac{1}{(4\pi)^3} \int F_i(\Omega)F_j(\Omega) \,\mathrm{d}\Omega.
\end{equation} 
$\mathbf{D}=\mathrm{diag}\{1, 1/3, 1/5, 1/15, 2/45, 2/45\}$ is diagonal due to the orthogonality of the terms $F_i(\Omega)$.

The $T(\Omega',\Omega)$ is subject to the detector effects (the limited acceptance of the detector, the detection and reconstruction efficiencies, and the resolution of the angular variables) that could affect the measured average of $F_i$. 
Figure~\ref{fig_mc_acc_eff} shows the detector effects in the distribution of some angular variables for the MC events.
At the level of MC generator level, without any simulation of the detector effects, the shown variables are uniformly distributed. Therefore, any structure observed in the distributions is due to detector effects. To illustrate the sensitivity, additional distributions in this figure show the same MC events re-weighted by three different PDFs with the values of the parameters as given in the figure caption. 
\begin{figure*}[tb]
 \centering
\includegraphics[width=0.49\textwidth]{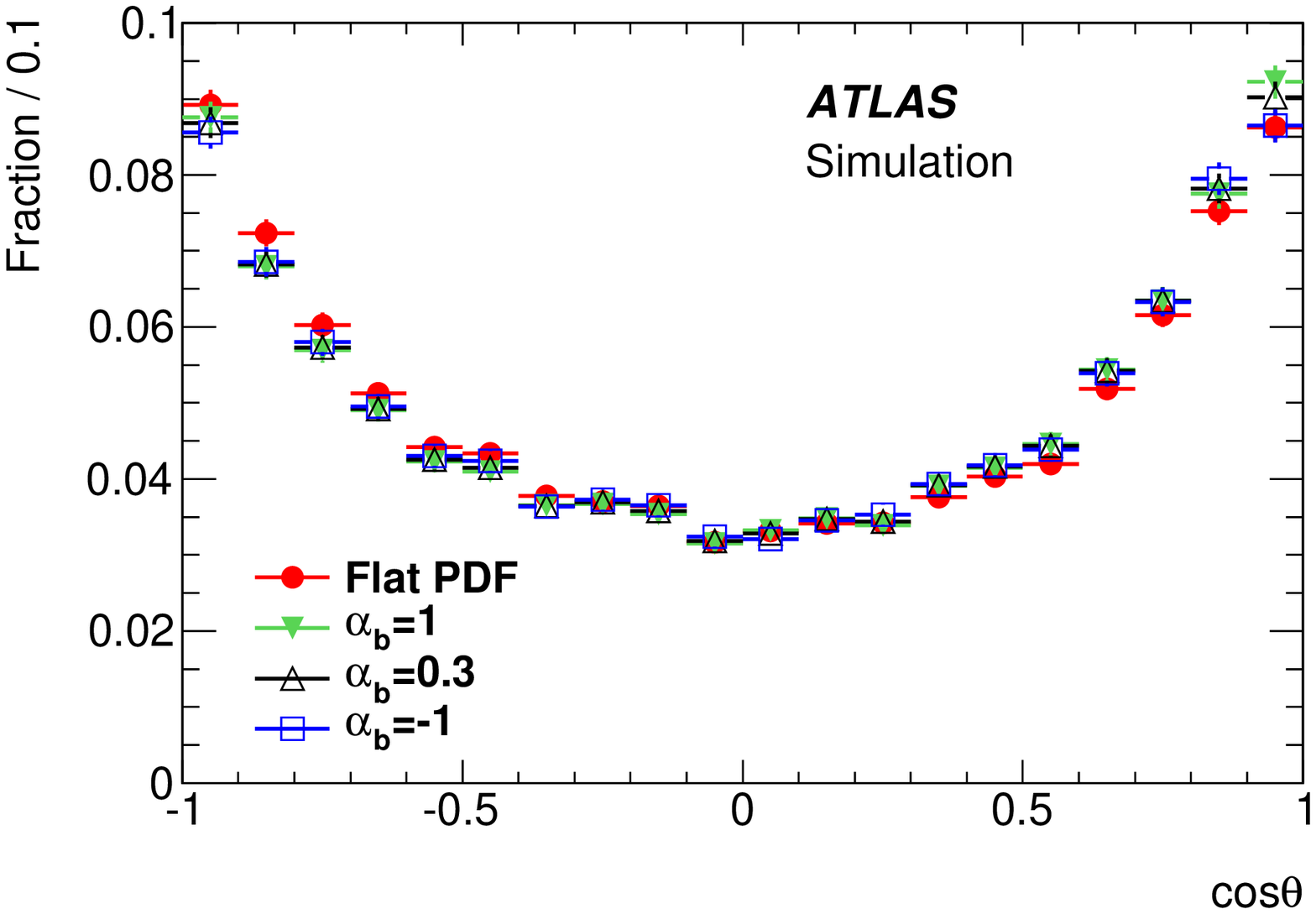}
\includegraphics[width=0.49\textwidth]{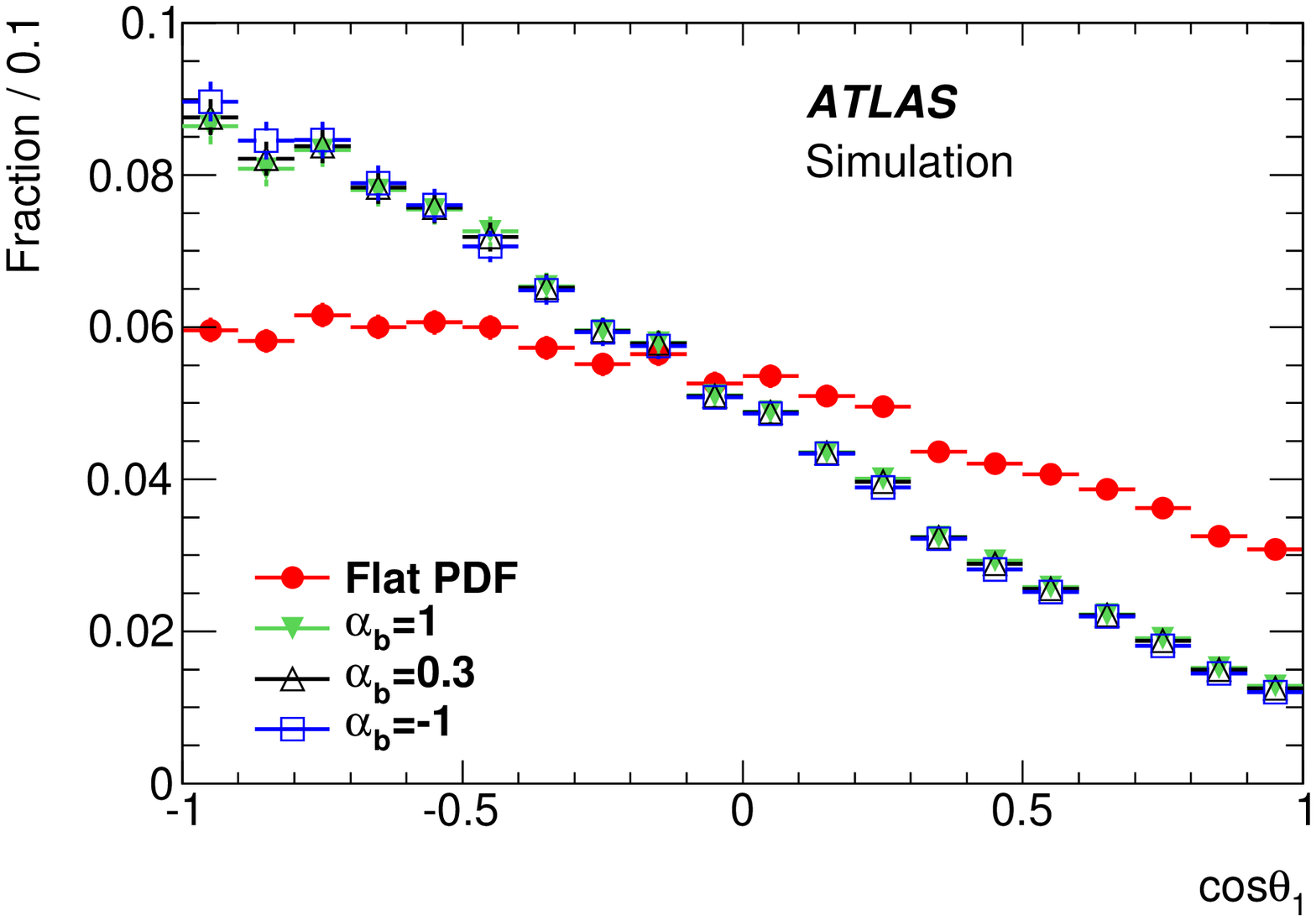}
\includegraphics[width=0.49\textwidth]{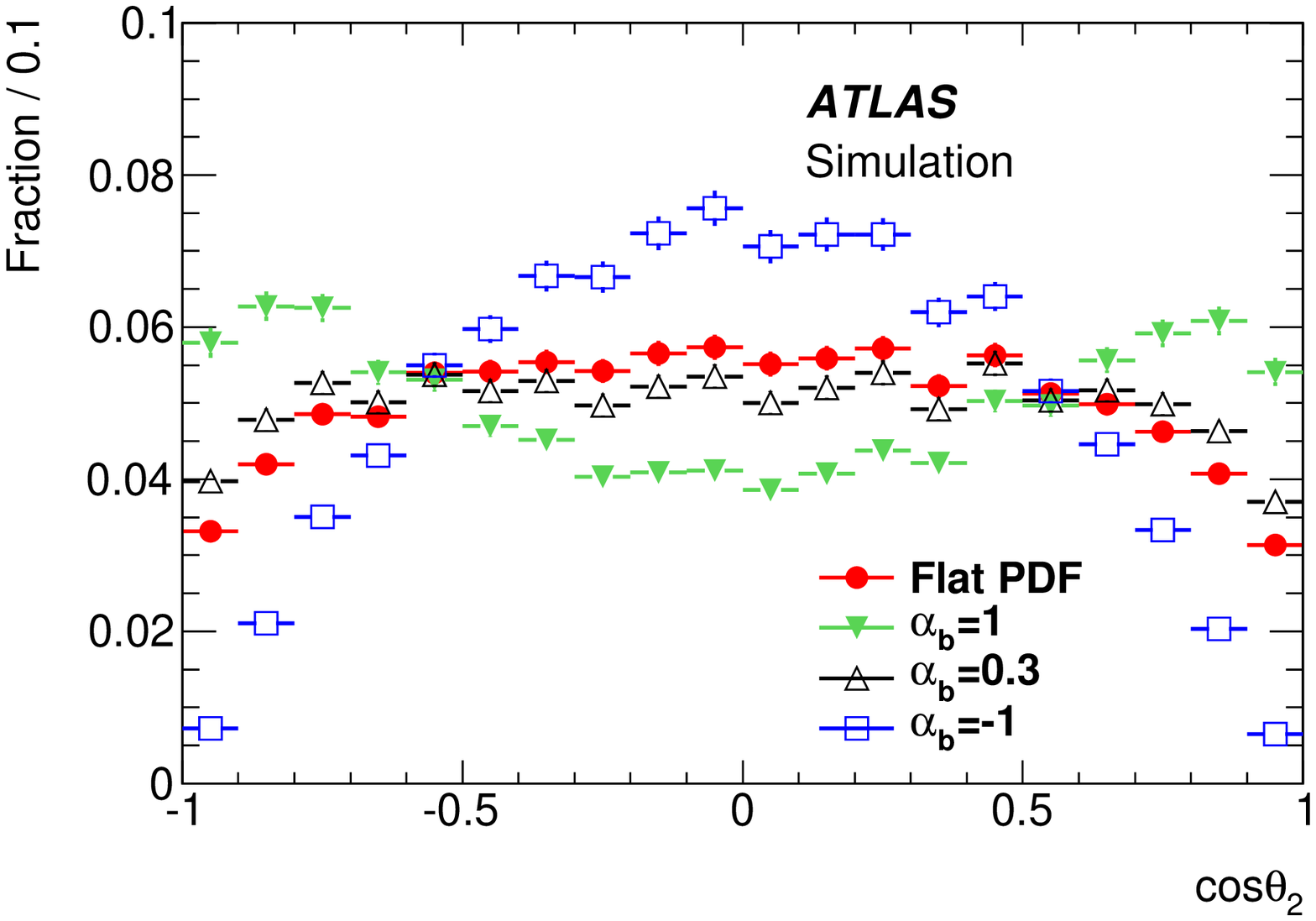}
\includegraphics[width=0.49\textwidth]{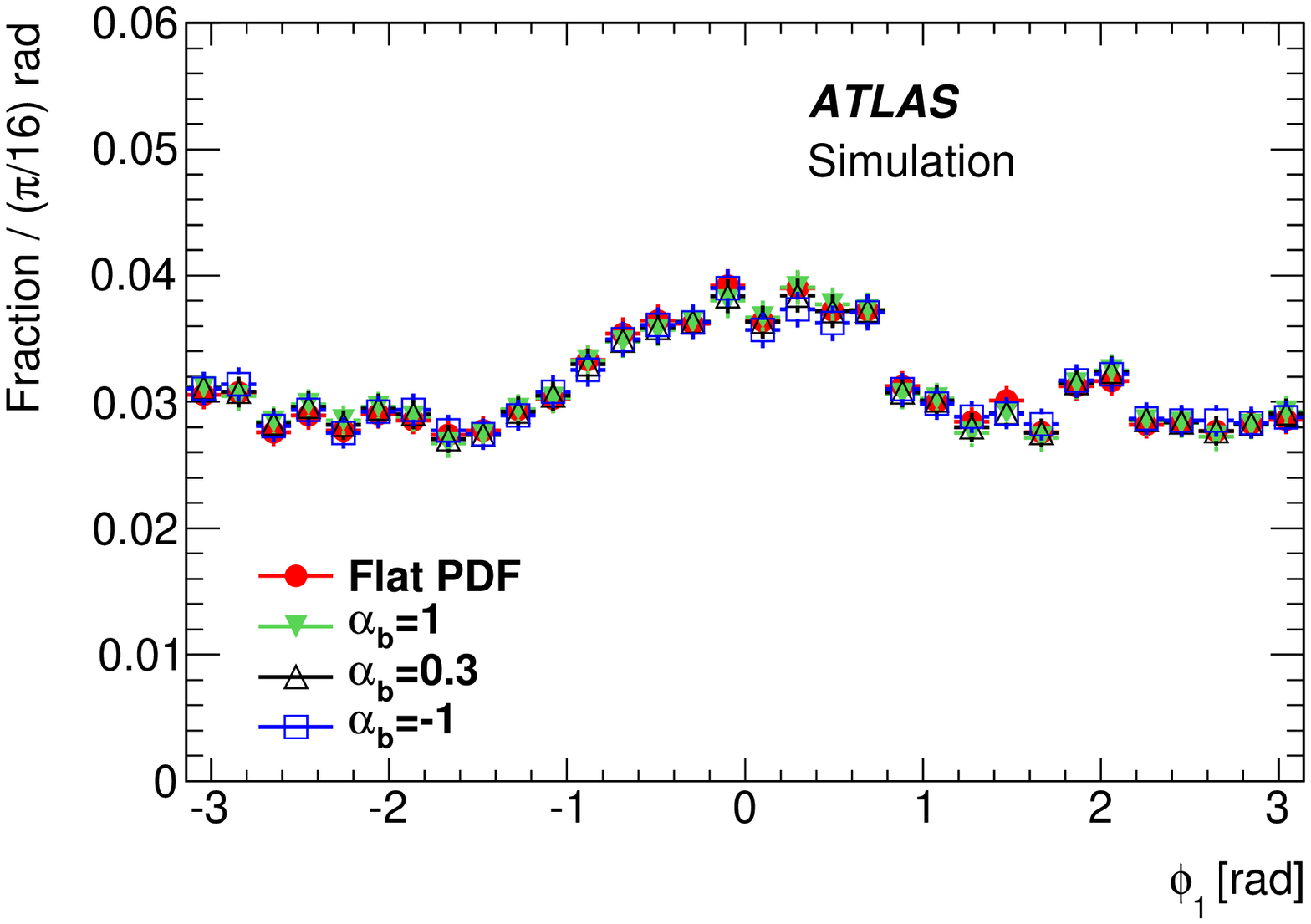}
\includegraphics[width=0.49\textwidth]{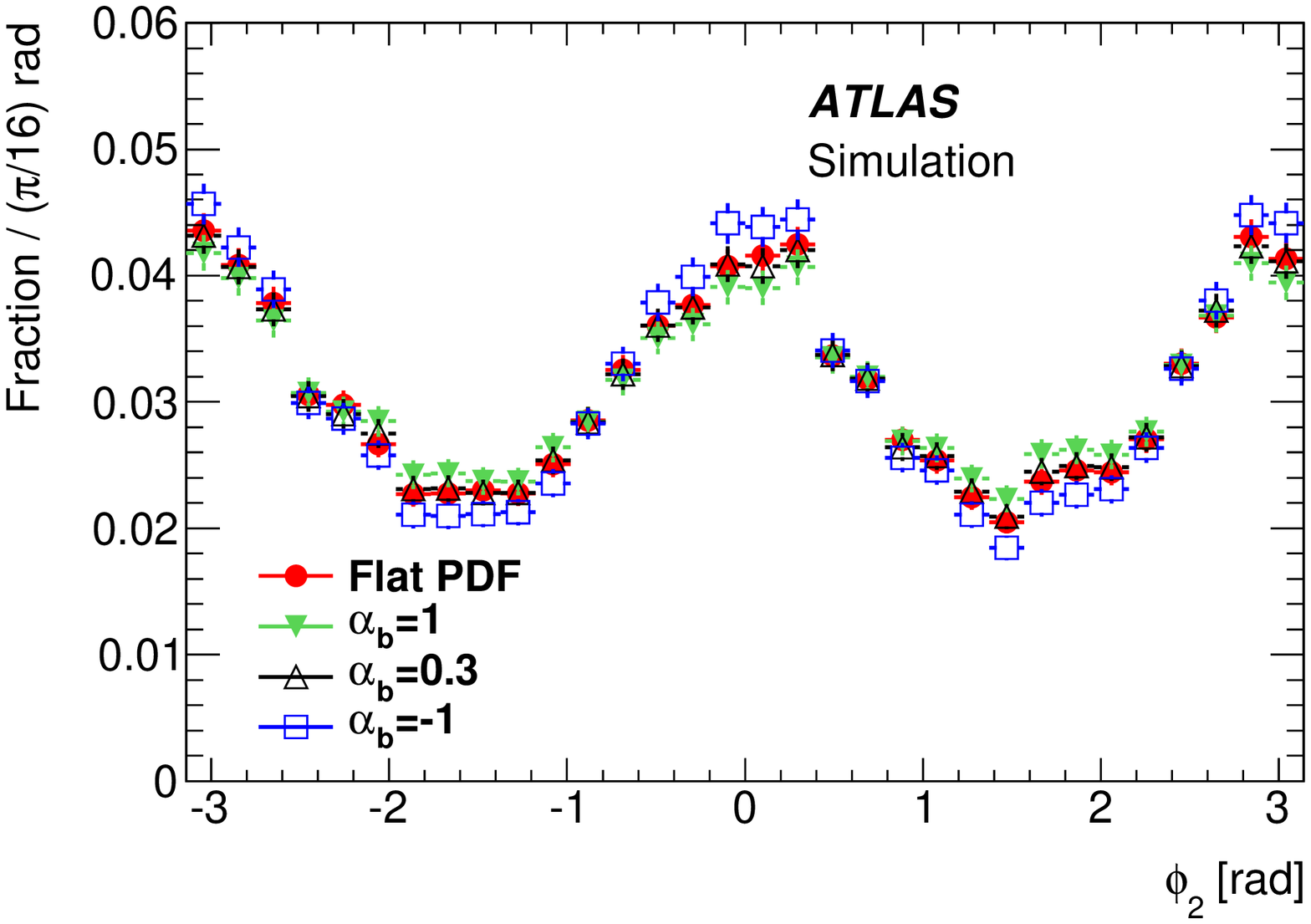}
\includegraphics[width=0.49\textwidth]{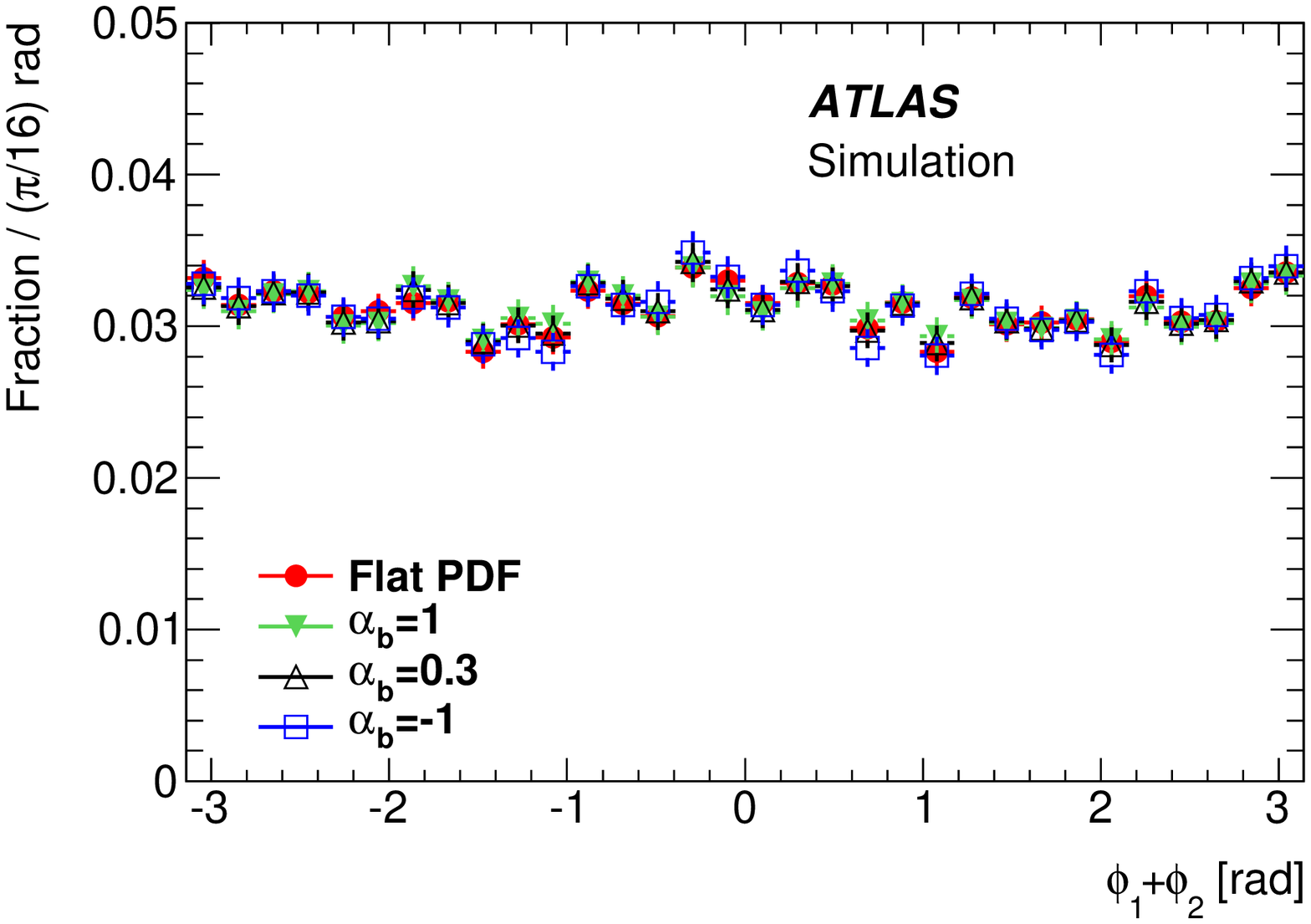}
\caption{Event distribution for each angular variable in simulated data after acceptance, efficiency and 
resolution effects are taken into account. The red filled points show the distributions in the default MC sample, where the generated distributions are uniform in all angular variables. For illustration of the sensitivity, the default MC events weighted using PDFs with $\alphab=\pm1$ (green filled down triangles and blue open squares) and the measured value $\alphab=0.3$ (open up triangles) are also shown.
Other parameters are set to $k_+=0.21$ and $k_-=0.13$ (measured values), and $\Delta_+=\Delta_-=0$.}
\label{fig_mc_acc_eff}
\end{figure*}

As shown in the preceding section, the matrix $\mathbf{C}$ is independent of the helicity amplitude parameters and can therefore be estimated using MC simulation, provided the detector is correctly described. For every reconstructed MC event, values of the true and reconstructed decay angles, $\Omega$ and $\Omega'$, are known. Their PDF can be written as
\begin{equation}
  w^{\rm mc}(\Omega', \Omega) = \frac{1}{\epsilon_{\rm T}} T(\Omega', \Omega) w^{\rm gen}(\Omega),
\end{equation}
where $w^{\rm gen}(\Omega)$ is the generator-level PDF and $\epsilon_{\rm T}$ is the overall normalization factor. Since a uniform angular distribution is used to generate the MC sample, $w^{\rm gen}(\Omega) = 1$, the distribution of angles $\Omega'$ and $\Omega$ for the reconstructed events is given solely by the detector effects. Therefore, the function $T(\Omega', \Omega)$ is also the PDF 
for the reconstructed MC events (except for the overall normalization factor $\epsilon_{\rm T}$), and Eq.~(\ref{eqn:C_matrix}) becomes a calculation of the mean of the expression $F_i(\Omega')F_j(\Omega)$ for variables $\Omega'$ and $\Omega$ distributed according to $T(\Omega', \Omega)$. The MC integration method is used to estimate the value of the coefficients $C_{ij}$ by replacing the integral with a summation:
\begin{align}
C_{ij} &= \frac{1}{(4\pi)^3}\iint F_i(\Omega') F_j(\Omega) T(\Omega', \Omega) \,\mathrm{d}\Omega' \mathrm{d}\Omega \nonumber \\
       &= \frac{\epsilon_{\rm T}}{(4\pi)^3}\iint F_i(\Omega') F_j(\Omega)\, w^{\rm mc}(\Omega', \Omega) \,\mathrm{d}\Omega' \mathrm{d}\Omega  \nonumber \\
       &\approx \frac{\epsilon_{\rm T}}{N^{\rm mc}} \sum_n^{N^{\rm mc}} F_i(\Omega'_n) F_j(\Omega_n).
\label{eqn:C_mc}
\end{align}
The unknown normalization factor, $\epsilon_{\rm T}$, can be determined from the constraint $\langle F_0\rangle^{\rm expected} \equiv 1$. The MC events used in the matrix $\mathbf{C}$ calculation are required to satisfy the same selection criteria as data. 
In order to have the same kinematics as data, two types of weights are applied to the MC events.
The first type is used to reproduce the same trigger configuration. The second one is used to reproduce the measured $(\pt, \eta)$ distribution of $\lamB$ candidates. The latter weight is called the kinematic weight and it is derived by comparing the two-dimensional $15\times10$ binned $(\pt, \eta)$ distribution of $\lamB$ in MC simulation and sideband-subtracted data.

The matrix $\mathbf{C}$ used in this analysis after weighting is:
\begin{equation}
\footnotesize
\begin{pmatrix*}[l]
\phantom{-}1  &  -0.113  &  -0.033  &  \phantom{-}0.0074  &  \phantom{-}0.0223  &  -0.0028 \\
-0.112  &  \phantom{-}0.3091  &  \phantom{-}0.0071  &  -0.0133  &  \phantom{-}0.0029  &  -0.0010 \\
-0.033  &  \phantom{-}0.0074  &  \phantom{-}0.1775  &  -0.0186  &  \phantom{-}0.0041  &  -0.0001 \\
\phantom{-}0.0071  &  -0.0133  &  -0.0185  &  \phantom{-}0.0545  &  \phantom{-}0.00013  &  \phantom{-}0.00029 \\
\phantom{-}0.0221  &  \phantom{-}0.0026  &  \phantom{-}0.0040  &  \phantom{-}0.00015  &  \phantom{-}0.0465  &  \phantom{-}0.0005 \\
-0.0031  &  -0.0008  &  -0.0003  &  \phantom{-}0.00034  &  \phantom{-}0.0005  &  \phantom{-}0.0450
\end{pmatrix*}.
\end{equation}
The MC statistical uncertainty of the elements on the diagonal is about 1\%, while the relative uncertainty of some of the off-diagonal elements is larger due to their small value. The impact of these uncertainties is discussed in Sec.~\ref{sec_syst}.

\subsection{Fit results}

The measured values of the averages $\langle F_i \rangle$ after the sideband subtraction and \Bd\ background correction are
\begin{align}
\langle F_2 \rangle &=-0.282\pm0.021, \nonumber \\
\langle F_4 \rangle &=-0.044\pm0.017, \nonumber \\
\langle F_6 \rangle &=\phantom{-}0.001\pm0.010, \nonumber \\
\langle F_{18} \rangle &=\phantom{-}0.019\pm0.009, \nonumber \\
\langle F_{19} \rangle &=-0.002\pm0.009.
\end{align}
The correlations between these measurements are listed in Table~\ref{tab:correlation_F}. In general, the correlations are small, except for the correlation of $\langle F_4 \rangle$ and $\langle F_6 \rangle$.
\begin{table}[tb]
 \center
\caption{Correlation matrix of the $\langle F_i \rangle$ measurements.}
\label{tab:correlation_F}
\begin{tabular}{llllll}
\hline \hline
$\langle F_i \rangle$        & $\langle F_2 \rangle$ & $\langle F_4 \rangle$     & $ \langle F_6 \rangle$     & $\langle F_{18} \rangle$  & $\langle F_{19} \rangle$  \\
 \hline
 $\langle F_2 \rangle$       & 1  &       $-0.066$ &       $-0.121$ & $\phantom{-}0.028$ & $\phantom{-}0.003$\\
 $\langle F_4 \rangle$       &    & $\phantom{-}1$ &       $-0.503$ & $\phantom{-}0.088$ & $\phantom{-}0.000$\\
 $\langle F_6 \rangle$       &    &                & $\phantom{-}1$ &           $-0.025$ &     $-0.008$\\
 $\langle F_{18} \rangle$    &    &                &                &     $\phantom{-}1$ & $\phantom{-}0.048$\\
 $\langle F_{19} \rangle$    &    &                &                &                    & $\phantom{-}1$ \\ 
\hline \hline
\end{tabular}
\end{table}

\begin{figure}[tb]
\centering
 \includegraphics[width=0.49\textwidth]{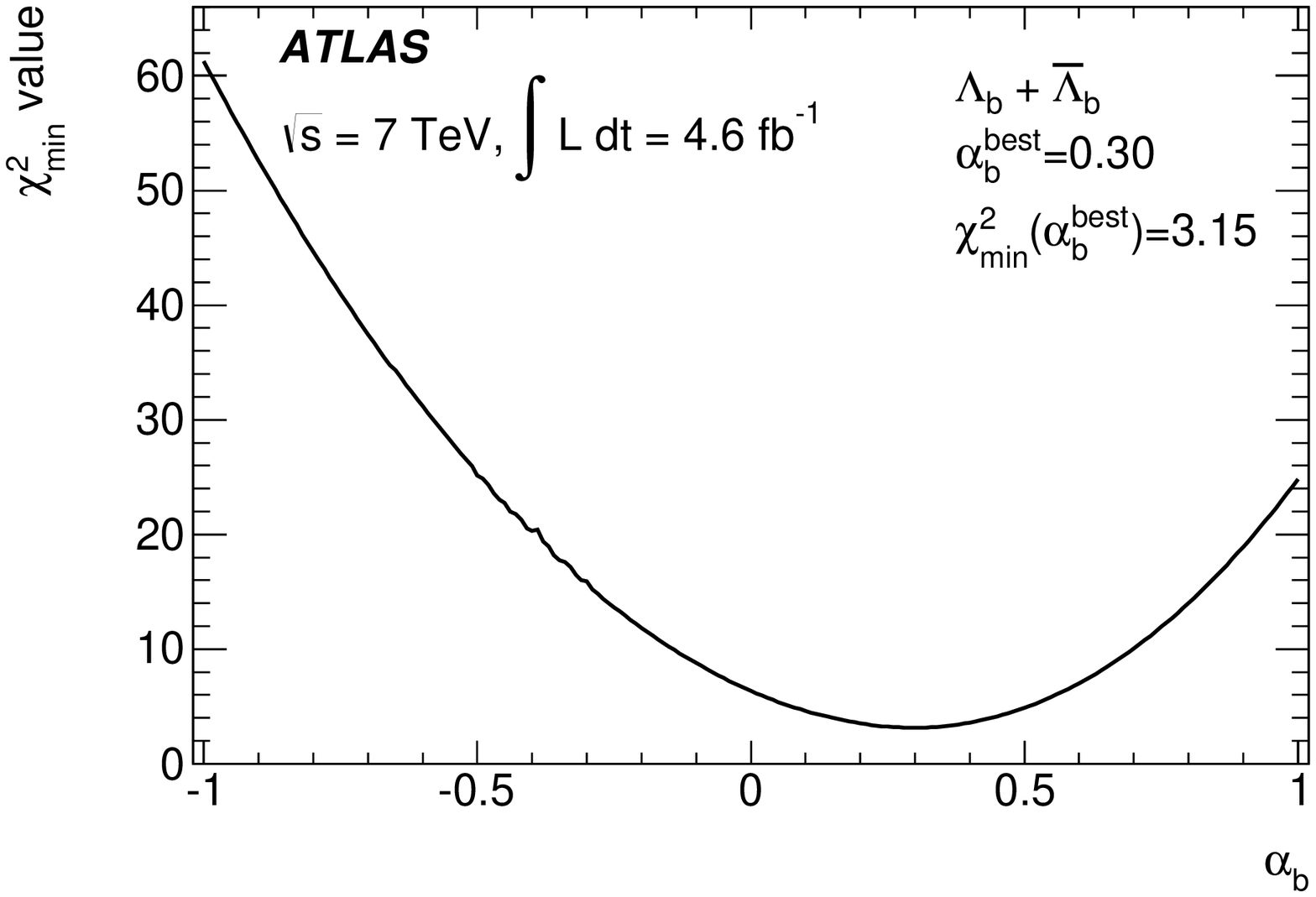}
\caption{The conditional $\chi^2_\mathrm{min}$ as a function of $\alphab$.}
\label{fig_alphab_scan}
\end{figure}

The $\chi^2$ fit (Eq.~(\ref{eqn_chi2})) is applied to data and yields:
\begin{align}
\alphab &= 0.30\pm0.16,\nonumber \\
k_+ &= 0.21_{-0.21}^{+0.14}, \nonumber \\
k_- &= 0.13_{-0.13}^{+0.20}. 
\end{align}
The statistical uncertainty of the parameters are calculated by finding the range that satisfies $\chi^2-\chi^2_{\rm min}<1$.
Negative values of $k_+$ and $k_-$ are allowed but they will give identical $\chi^2$, because the real values used in fit are $|k_+|$ and $|k_-|$. Thus, negative-value parts of their uncertainty bands are truncated.
With the limited data sample size, values of the relative phases, $\Delta_+$ and $\Delta_-$, obtained from the fit, are consistent with the entire allowed range, $[-\pi,\pi]$. 
The effect of their large uncertainties on the determination of $\alphab$, $k_+$ and $k_-$ is checked in an alternative fit.
Since the phase parameters are not well determined, and the efficiency of the measurement does not have a strong dependence on $\phi_1+\phi_2$ as shown in Fig.~\ref{fig_mc_acc_eff}, only the first four terms in Table~\ref{tab:pdf2} are considered in the alternative fit and only the parameters $\alphab$, $k_+$ and $k_-$ are determined. 
The results of this fit, both the central values and the statistical uncertainties, are very similar to those of the main analysis. In particular, the differences between the central values are smaller than the statistical errors and comparable to the systematic uncertainties discussed in Sec.~\ref{sec_syst}.
Figure~\ref{fig_alphab_scan} shows the $\chi^2_\mathrm{min}$ as a function of the assumed $\alphab$ parameter with the condition that the $\alphab$ parameter is fixed in the fit. The minimum of this conditional $\chi^2_\mathrm{min}$ curve gives the central value of $\alphab$ $(\alphab^{\rm best})$ and the corresponding $\chi^2$ value is 3.15.
The correlation matrix of the fitted parameters is shown in Table~\ref{tab:correlation_p}. 
There are no strong correlations between these parameters.
The corresponding absolute values of the helicity amplitudes are
\begin{align}
|a_{+}| &=0.17_{-0.17}^{+0.12}, \nonumber \\
|a_{-}| &=0.59_{-0.07}^{+0.06}, \nonumber \\
|b_{+}| &=0.79_{-0.05}^{+0.04}, \nonumber \\
|b_{-}| &=0.08_{-0.08}^{+0.13}.
\end{align}

\begin{figure*}[tb]
 \centering
 \includegraphics[width=0.49\textwidth]{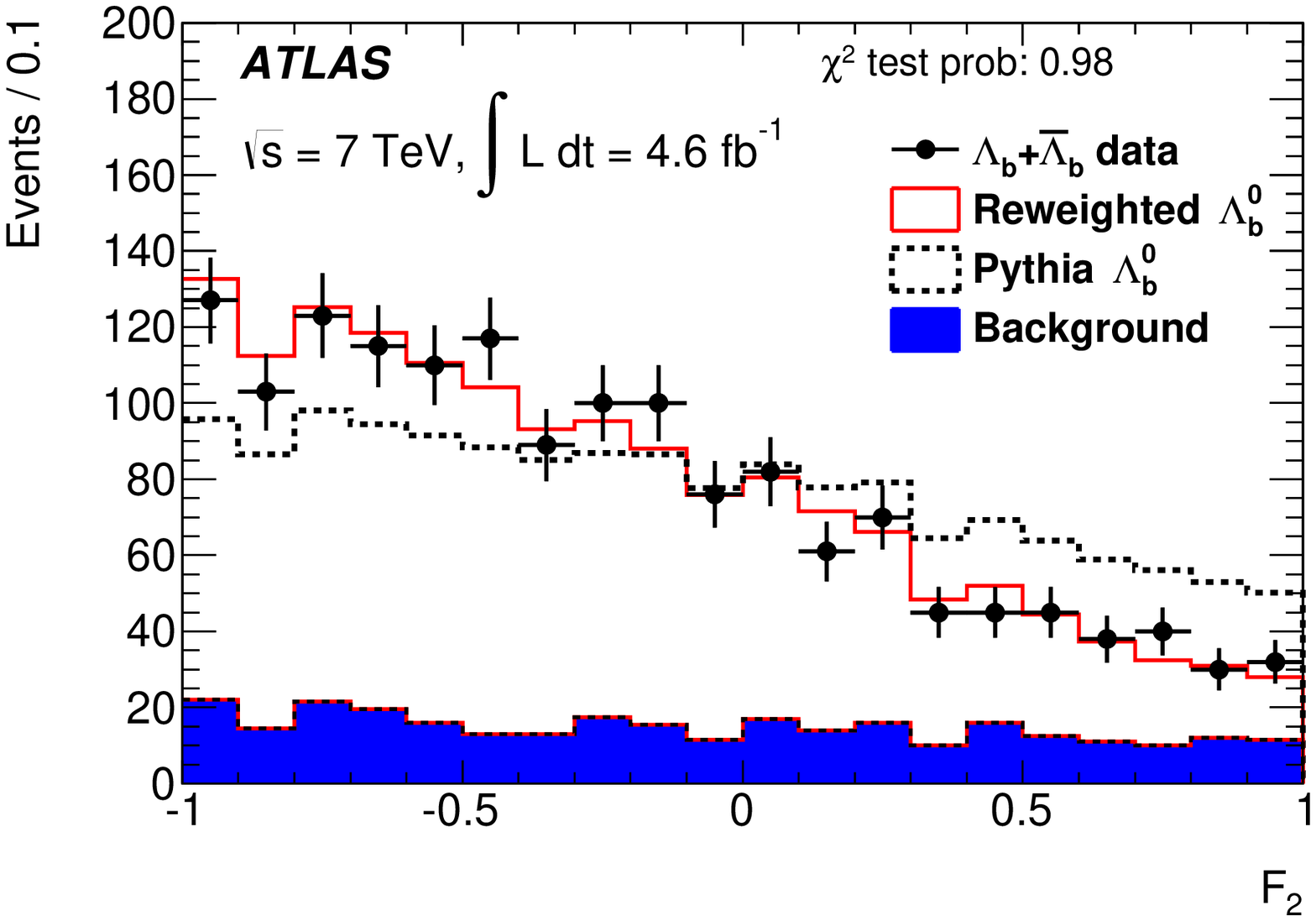}
 \includegraphics[width=0.49\textwidth]{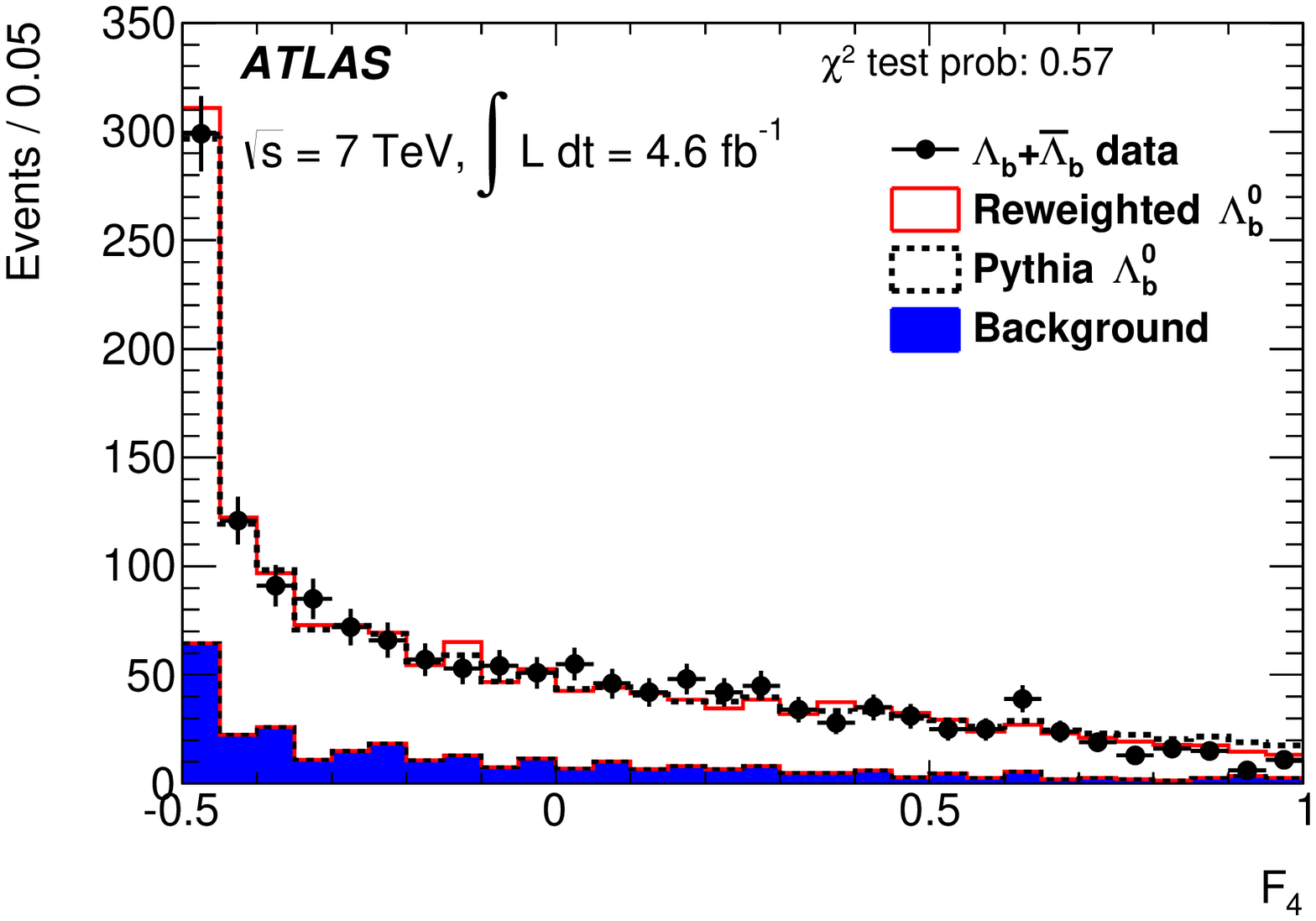}
 \includegraphics[width=0.49\textwidth]{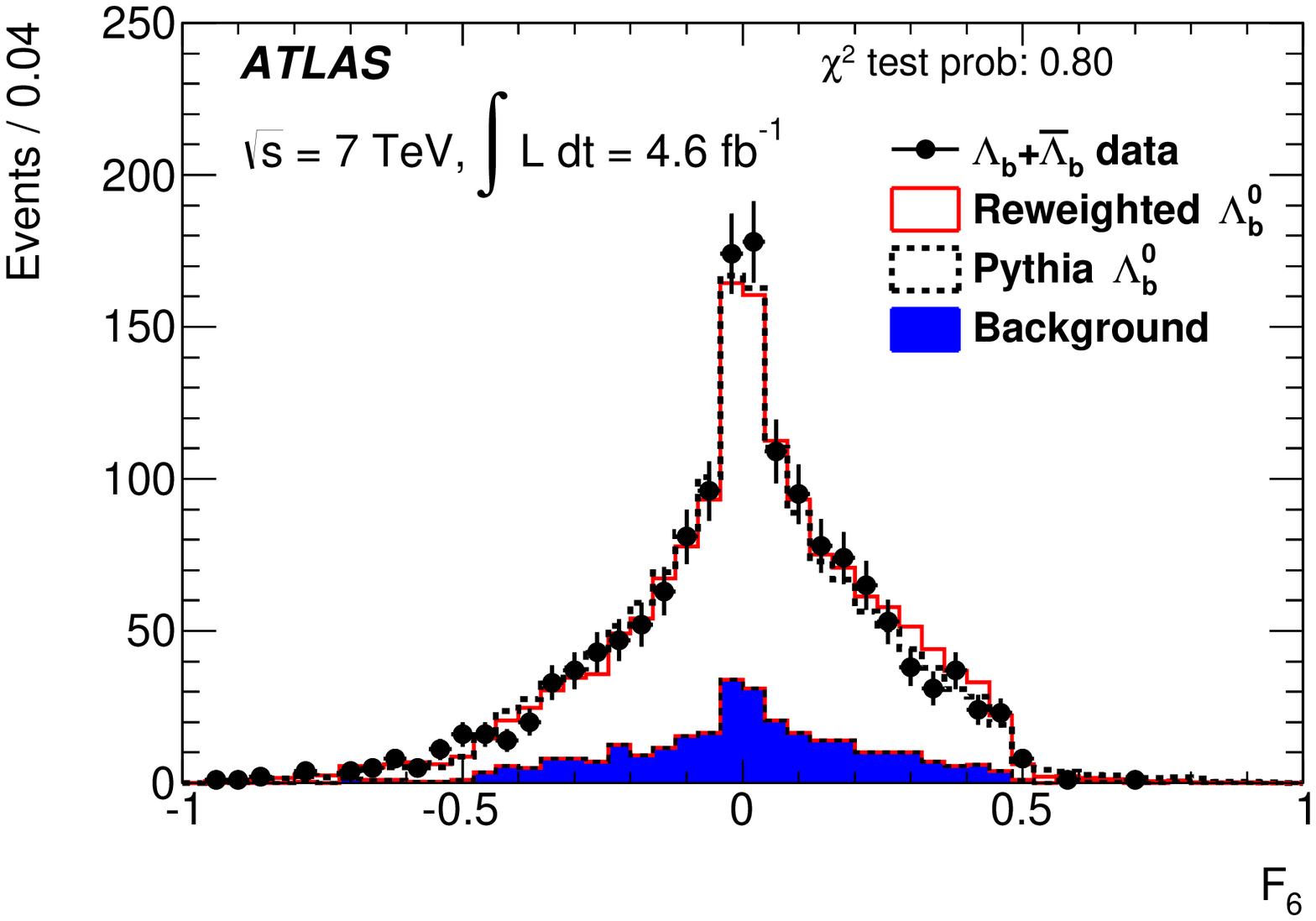}
 \includegraphics[width=0.49\textwidth]{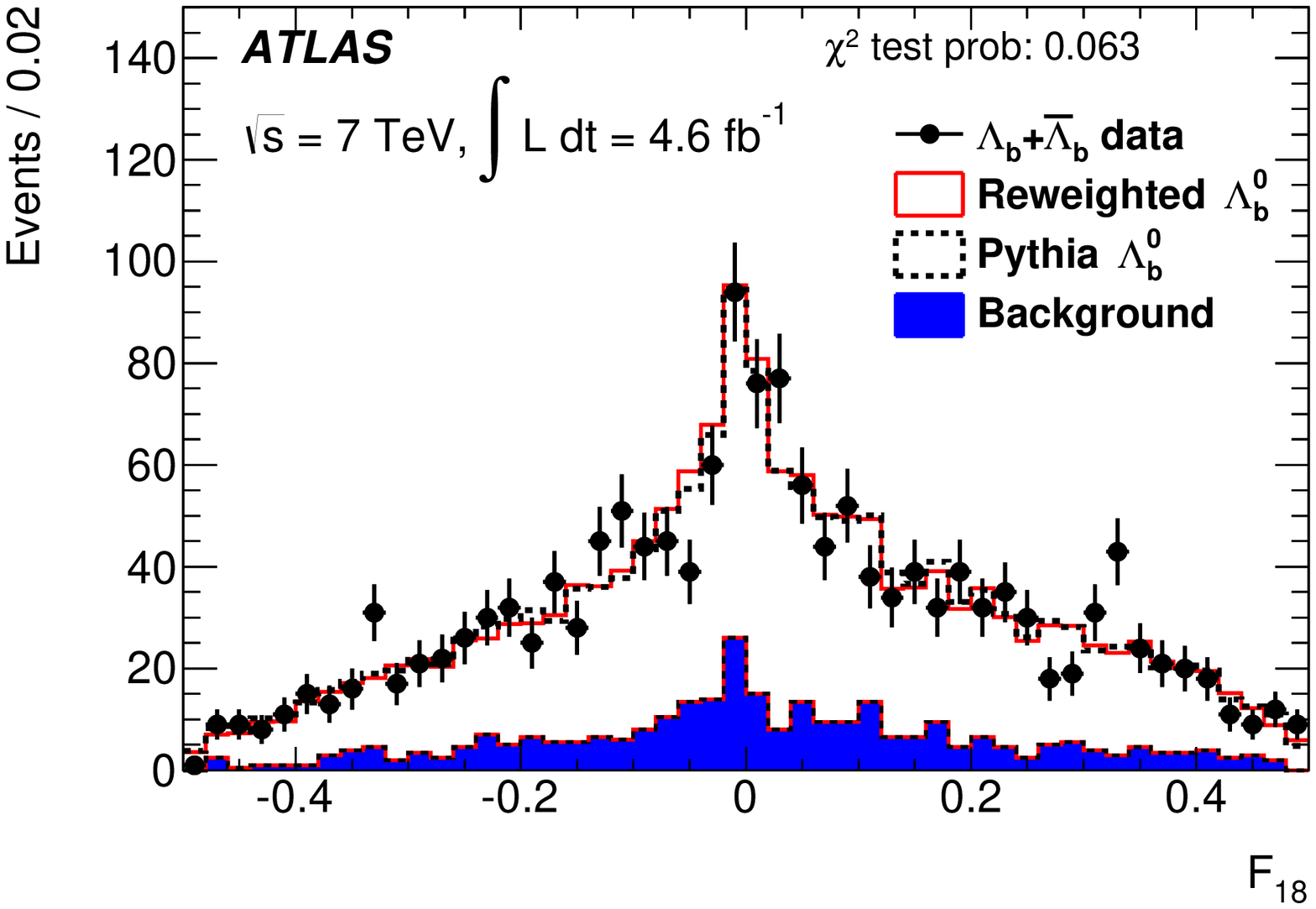}
 \includegraphics[width=0.49\textwidth]{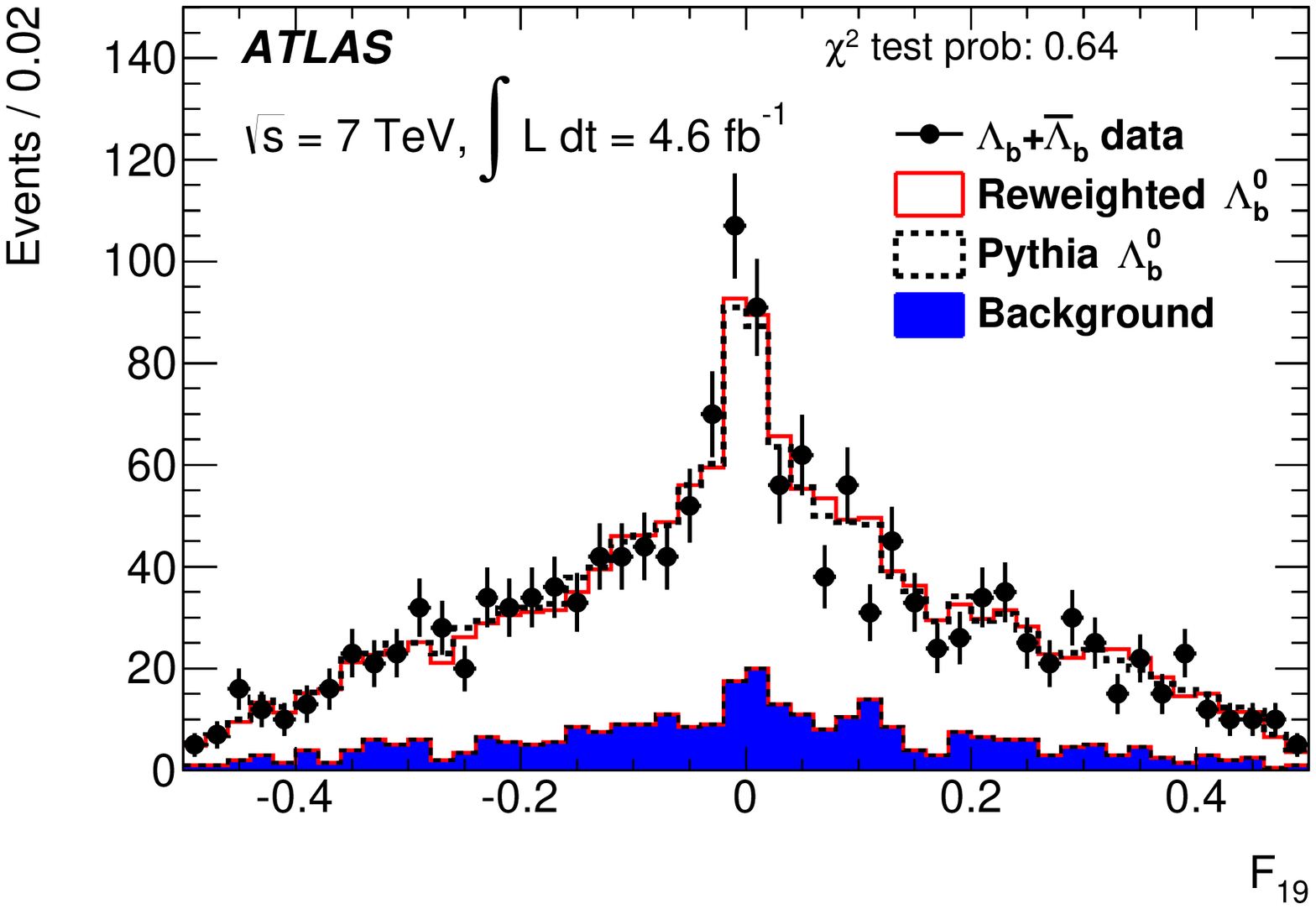}
\caption{The predicted distributions of $F_i$ variables from the sum of the weighted MC events (red line) and the background (blue area) are compared with data (black points). The background is estimated by adding the left and right sidebands and scaling by 0.5. The $\chi^2$-test probability of each comparison is shown in the top left corner of the plot. The predictions of the unweighted MC events (black dashed line) are also shown.}
\label{fig_fit_res_check1}
\end{figure*}

To check the fit results, the MC events are further weighted using the signal PDF with parameters determined from the fit and normalized to the number of events of the sideband-subtracted data. These weighted MC events and sideband background distributions of $F_i$ are added and compared with data. Figure~\ref{fig_fit_res_check1} shows good agreement between the weighted MC events and data.

\begin{table}[tb]
 \centering
\caption{Correlation matrix of the fitted parameters.}
\label{tab:correlation_p}
\begin{tabular}{llll}
\hline \hline
Parameter  & $\alphab$ & $k_+$ & $k_-$  \\
\hline
$\alphab$ & 1 & $0.41$ & $-0.19 $  \\
$k_+$      &   & $1   $ & $\phantom{-}0.20$  \\
$k_-$      &   &        & $\phantom{-}1    $  \\
\hline \hline
\end{tabular}
\end{table}

The polarization of $\lamB$ and $\lamBBar$ is checked with data and is found consistent with the expected value of zero (Sec.~\ref{sec:theory}). The combination of \LamB\ and \LamBBar\ samples is also justified by the consistent results from the separate fits for the two samples.
\section{Systematic uncertainties}
\label{sec_syst}

\begin{table*}[tb]
 \centering
\caption{Systematic uncertainties.}
\label{tab_systematics}
\begin{tabular}{llllllll}
\hline\hline
Source & $\alphab$ & $k_+$ & $k_-$ & $|a_+|$ & $|a_-|$ & $|b_+|$ & $|b_-|$\\ \hline
Background shape & 0.034 & 0.020 & 0.042 & 0.018 & 0.017 & 0.010 & 0.024 \\
\Bd\ background & 0.011 & 0.085 & 0.061 & 0.069 & 0.008 & 0.008 & 0.036 \\
Angles resolution & 0.005 & 0.017 & 0.026 & 0.014 & 0.004 & 0.002 & 0.015 \\
MC mass resolution modeling & 0.020 & 0.004 & 0.004 & 0.002 & 0.008 & 0.007 & 0.002 \\
MC kin. weighting (MC parameterization) & 0.007 & 0.010 & 0.008 & 0.008 & 0.007 & 0.002 & 0.005 \\
MC kin. weighting (data sample size) & 0.011 & 0.017 & 0.014 & 0.014 & 0.005 & 0.003 & 0.008 \\
MC sample size & 0.047 & 0.090 & 0.121 & 0.039 & 0.016 & 0.013 & 0.037 \\
Value of $\alphaL$ & 0.009 & 0.023 & 0.023 & 0.019 & 0.005 & 0.001 & 0.014 \\
\hline 
Total & 0.064 & 0.130 & 0.147 & 0.086 & 0.028 & 0.020 & 0.061\\ \hline \hline
\end{tabular}
\end{table*}

The systematic uncertainty in this measurement mainly comes from two sources: the measurement of the $\langle F_i \rangle$ moments and the calculation of the matrix $\mathbf{C}$. The systematic uncertainties considered in this analysis are listed below. The first two items refer to the first category, and the other items are related to the calculation of the matrix $\mathbf{C}$ and other uncertainties:
\begin{itemize}
  \item The shape of background. The effect of a possible non-linearity of the combinatorial background is checked by using the left or right sideband separately, instead of the average of the two sidebands, to estimate the background contribution in the central region. This gives a maximum difference of 0.034 in the $\alphab$ value.

  \item The \Bd\ background estimation. The number of \Bd\ background candidates is varied by one standard deviation. The impact of this variation on the $\alphab$ value is 0.011.

  \item The resolution of decay angles. The effect of decay angles' measurement resolution is accounted for by the matrix $\mathbf{C}$; however, it relies on the MC simulation. An uncertainty due to the angular resolution is conservatively estimated by replacing the generator-level decay angles with the reconstructed ones (and vice versa) in the matrix $\mathbf{C}$ calculation. The effect on $\alphab$ is found to be 0.005.

  \item The modeling of the mass resolution. 
The mass resolution scale factor is found to be $0.99\pm0.06$ by fitting the MC simulation to data. The scale factor in the MC simulation used in the matrix $\mathbf{C}$ calculation is varied from 0.93 to 1.05 to study the effect of possible mis-modeling. The maximum of the deviation from the nominal $\alphab$ is 0.020 and is taken as a systematic uncertainty.

  \item MC kinematic weight calculation uncertainty due to helicity parameters in MC simulation. The kinematic weight of each MC event is calculated by comparing the distributions of $(\pt, \eta)$ in the MC sample and background-subtracted data. The distribution of $(\pt, \eta)$ in the MC sample may slightly depend on values of the helicity amplitudes used in the MC production. To study this effect, the helicity parameters are varied and the fit is repeated using the new kinematic weights. The $\alphab$ parameter is varied from $-1$ to 1 and $k_+$, $k_-$ parameters are varied from 0 to 1.
The maximum change in $\alphab$ caused by this variation is 0.007, and this is taken as a systematic uncertainty.
  \item MC kinematic weight calculation uncertainty due to limited data sample size. The effect of the limited data sample size in the kinematic weight calculation is 
estimated by varying the number of data events in each $(\pt, \eta)$ bin in the kinematic weight calculation. In each variation,
Poisson samplings of the numbers of data events in the signal region and in sidebands are used instead of the numbers themselves in each $(\pt, \eta)$ bin. The test is repeated 2000 times and the root mean square of the fit results is considered as a systematic uncertainty. The resulting uncertainty on $\alphab$ is 0.011.

  \item MC statistics. The statistical uncertainty of the measured moments, $\langle F_i \rangle$, is contained in the covariance matrix $\mathbf{V}$ in Eq.~(\ref{eqn_chi2}). However, this matrix does not contain the statistical uncertainty of the expected moments, $\langle F_i \rangle^{\rm expected}$, which arises from the limited MC sample size in the matrix $\mathbf{C}$ calculation. In order to parameterize the effect of this uncertainty, the covariance matrix $\mathbf{V}^{\rm MC}$ of the $\langle F_i \rangle^{\rm expected}$ moments is calculated using the MC events and is added to the covariance matrix in Eq.~(\ref{eqn_chi2}). The fit is repeated and the new uncertainties in the fitted parameters are estimated, this time including the uncertainty from both the data and MC sample statistics. The default values of the statistical uncertainties are subtracted in quadrature from the new ones to isolate the contribution of the limited MC sample size. In case of the $\alphab$ parameter, this uncertainty is estimated to be 0.047.
 
 \item The value of $\alphaL$, taken from Ref.~\cite{PDG}, is varied by one standard deviation to check the effect on the extracted parameters. The differences are taken as a systematic uncertainty, which is 0.009 for the value of $\alphab$.
\end{itemize}
The contributions of these sources to the systematic uncertainties of the measured parameters are summarized in Table~\ref{tab_systematics}.
The total systematic uncertainty is calculated by adding individual contributions in quadrature. The total uncertainty for $\alphab$ is 0.064.
\section{Conclusions}
\label{sec_con}
A measurement of the parity-violating decay asymmetry parameter $\alphab$ and the helicity amplitudes for the decay \LamBDecay\ has been performed using the 4.6~$\ifb$ $pp$ collisions at a center-of-mass energy of 7~TeV recorded by the ATLAS detector at the LHC in 2011. The measured values of $\alphab$, $k_+$ and $k_-$ are:
\begin{align}
\alphab &= 0.30 \pm 0.16 ({\rm stat})\pm 0.06 ({\rm syst}), \nonumber \\
k_+ &= 0.21_{-0.21}^{+0.14}({\rm stat})\pm 0.13 ({\rm syst}), \nonumber \\
k_- &= 0.13_{-0.13}^{+0.20}({\rm stat})\pm 0.15 ({\rm syst}),
\end{align}
corresponding to the value of helicity parameters
\begin{align}
|a_{+}|&=0.17^{+0.12}_{-0.17} ({\rm stat})\pm0.09 ({\rm syst}), \nonumber \\
|a_{-}|&=0.59^{+0.06}_{-0.07} ({\rm stat})\pm0.03 ({\rm syst}), \nonumber \\
|b_{+}|&=0.79^{+0.04}_{-0.05} ({\rm stat})\pm0.02 ({\rm syst}), \nonumber \\
|b_{-}|&=0.08^{+0.13}_{-0.08} ({\rm stat})\pm0.06 ({\rm syst}).
\end{align}
The $\lamB$ decay has large amplitudes $|a_-|$ and $|b_+|$, which means the negative-helicity states for \Lam\ are preferred.
The $\lam$ and $\jpsi$ from $\lamB$ decay are highly polarized.
Adding in quadrature the statistical and systematic uncertainties, the observed value of $\alphab$ is consistent with the recent measurement $\alphab=0.05 \pm 0.17({\rm stat}) \pm 0.07({\rm syst})$ by LHCb~\cite{Aaij:2013oxa} at the level of one standard deviation.
However, it is not consistent with the expectation from pQCD~\cite{Chou:2001bn} ($\alphab$ in the range from $-0.17$ to $-0.14$),
and HQET~\cite{Leitner:2004dt, Ajaltouni:2004zu} ($\alphab=0.78$) at a level of about 2.6 and 2.8 standard deviations, respectively.
\section*{Acknowledgments}
We thank CERN for the very successful operation of the LHC, as well as the
support staff from our institutions without whom ATLAS could not be
operated efficiently.

We acknowledge the support of ANPCyT, Argentina; YerPhI, Armenia; ARC,
Australia; BMWF and FWF, Austria; ANAS, Azerbaijan; SSTC, Belarus; CNPq and FAPESP,
Brazil; NSERC, NRC and CFI, Canada; CERN; CONICYT, Chile; CAS, MOST and NSFC,
China; COLCIENCIAS, Colombia; MSMT CR, MPO CR and VSC CR, Czech Republic;
DNRF, DNSRC and Lundbeck Foundation, Denmark; EPLANET, ERC and NSRF, European Union;
IN2P3-CNRS, CEA-DSM/IRFU, France; GNSF, Georgia; BMBF, DFG, HGF, MPG and AvH
Foundation, Germany; GSRT and NSRF, Greece; ISF, MINERVA, GIF, I-CORE and Benoziyo Center,
Israel; INFN, Italy; MEXT and JSPS, Japan; CNRST, Morocco; FOM and NWO,
Netherlands; BRF and RCN, Norway; MNiSW and NCN, Poland; GRICES and FCT, Portugal; MNE/IFA, Romania; MES of Russia and ROSATOM, Russian Federation; JINR; MSTD,
Serbia; MSSR, Slovakia; ARRS and MIZ\v{S}, Slovenia; DST/NRF, South Africa;
MINECO, Spain; SRC and Wallenberg Foundation, Sweden; SER, SNSF and Cantons of
Bern and Geneva, Switzerland; NSC, Taiwan; TAEK, Turkey; STFC, the Royal
Society and Leverhulme Trust, United Kingdom; DOE and NSF, United States of
America.

The crucial computing support from all WLCG partners is acknowledged
gratefully, in particular from CERN and the ATLAS Tier-1 facilities at
TRIUMF (Canada), NDGF (Denmark, Norway, Sweden), CC-IN2P3 (France),
KIT/GridKA (Germany), INFN-CNAF (Italy), NL-T1 (Netherlands), PIC (Spain),
ASGC (Taiwan), RAL (UK) and BNL (USA) and in the Tier-2 facilities
worldwide.

\bibliographystyle{apsrev4-1a}
\bibliography{paper}

\onecolumngrid
\clearpage
\begin{flushleft}
{\Large The ATLAS Collaboration}

\bigskip

G.~Aad$^{\rm 84}$,
T.~Abajyan$^{\rm 21}$,
B.~Abbott$^{\rm 112}$,
J.~Abdallah$^{\rm 152}$,
S.~Abdel~Khalek$^{\rm 116}$,
O.~Abdinov$^{\rm 11}$,
R.~Aben$^{\rm 106}$,
B.~Abi$^{\rm 113}$,
M.~Abolins$^{\rm 89}$,
O.S.~AbouZeid$^{\rm 159}$
% ...
\bigskip
\\
% ...
$^{11}$ Institute of Physics, Azerbaijan Academy of Sciences, Baku, Azerbaijan\\
% ...
$^{21}$ Physikalisches Institut, University of Bonn, Bonn, Germany\\
% ...
$^{84}$ CPPM, Aix-Marseille Universit{\'e} and CNRS/IN2P3, Marseille, France\\
% ...
$^{89}$ Department of Physics and Astronomy, Michigan State University, East Lansing MI, United States of America\\
% ...
$^{106}$ Nikhef National Institute for Subatomic Physics and University of Amsterdam, Amsterdam, Netherlands\\
% ...
$^{112}$ Homer L. Dodge Department of Physics and Astronomy, University of Oklahoma, Norman OK, United States of America\\
$^{113}$ Department of Physics, Oklahoma State University, Stillwater OK, United States of America\\
% ...
$^{116}$ LAL, Universit{\'e} Paris-Sud and CNRS/IN2P3, Orsay, France\\
% ...
$^{152}$ Institute of Physics, Academia Sinica, Taipei, Taiwan\\
% ...
$^{159}$ Department of Physics, University of Toronto, Toronto ON, Canada\\
% ...
\end{flushleft}
%

\end{document}